\documentclass[letter,11pt]{article}
\pdfoutput=1
\usepackage{heppub} 
\allowdisplaybreaks
\usepackage{hyperref}
\usepackage{graphicx}
\usepackage{amsmath}
\usepackage{amssymb} 
\usepackage[normalem]{ulem}
\usepackage{slashed}
\usepackage[utf8]{inputenc}
\usepackage[T1]{fontenc}
\usepackage{subcaption}
\usepackage{scalerel,stackengine}
\usepackage{multirow}
\usepackage{array}
\newcommand{\parallelsum}{\mathbin{\!/\mkern-5mu/\!}}
\newcommand{\nn}{\nonumber}

\def\bea#1\eea{\begin{align}#1\end{align}}
\newcommand{\bs}{\boldsymbol}

\preprint{IQuS@UW-21-106}

\title{Quantum Simulation of QED in Coulomb Gauge}

\author{Xiaojun Yao}

\affiliation{InQubator for Quantum Simulation, Department of Physics, University of Washington, Seattle, WA 98195, USA}

\emailAdd{xjyao@uw.edu}

\abstract{A recent work considered quantum simulation of Quantum Electrodynamics on a lattice in the Coulomb gauge with gauge degrees of freedom represented in the occupation basis in momentum space. Here we consider the more efficient representation of the gauge degrees of freedom in field basis in position space and develop a quantum algorithm for real-time simulation. We show that the continuum Coulomb gauge Hamiltonian is equivalent to the temporal gauge Hamiltonian when acting on physical states consisting of fermion and transverse gauge fields. The Coulomb gauge Hamiltonian is discretized by using the Green's function of the discrete Laplacian operator under the Dirichlet boundary conditions.
Both the continuum Coulomb gauge Hamiltonian and the discretized one proposed here guarantee that the unphysical longitudinal gauge fields are decoupled and commute with the corresponding Hamiltonian. Thus there is no need to impose any constraint. The local gauge field basis and the canonically conjugate variable basis are swapped efficiently using the quantum Fourier transform. We prove that the qubit cost to represent physical states and the gate count for real-time simulation scale polynomially with the lattice size, energy, time, accuracy, and Hamiltonian parameters in lattice units. The gate cost here for implementing the time evolution of the gauge field is reduced at least by a factor on the order of $10^8$ for modest lattice size and accuracy level compared with the previous work.}

\begin{document}
\maketitle

\section{Introduction}
\label{sec:intro}
Quantum computing appeals to widespread interest as a potential nonperturbative tool to study field theories, which compose the underlying theoretical framework for the Standard Model that describes three of the four fundamental forces in nature~\cite{Zohar:2015hwa,Banuls:2019bmf,Klco:2021lap,Bauer:2022hpo,Bauer:2023qgm,DiMeglio:2023nsa,Beck:2023xhh}. In the case of 1+1 dimensions (1+1D), last few years have seen much significant progress that enables classical and quantum simulations for about $100$ qubits~\cite{Farrell:2023fgd,Farrell:2024fit,Lin:2024eiz,Farrell:2025nkx,Schuhmacher:2025ehh,Florio:2025hoc,Banuls:2025wiq}. Many studies now move on to higher dimensions~\cite{Osborne:2022jxq,Muller:2023nnk,Itou:2024psm,Mueller:2024mmk,Maiti:2024jwk,Illa:2025dou,Yang:2025edn,DiMarcantonio:2025cmf,Illa:2025njz,NSSrivatsa:2025jhh,Balaji:2025afl}.

For lattice scalar field theories, the pioneering work of Jordan, Lee, and Preskill (JLP) systematically estimated the quantum computing resources needed to simulate wave packets scattering, which scale polynomially with the number of particles, their energy, and the desired precision~\cite{Jordan:2011ci,Jordan:2012xnu}. Furthermore, they proved that scattering in scalar field theory falls into the computational complexity class Bounded-error Quantum Polynomial time (BQP)-complete~\cite{Jordan:2017lea}. It was shown later that this efficiency can be understood from the Nyquist-Shannon sampling theorem~\cite{Klco:2018zqz}.

For lattice gauge theories, most studies used the Kogut-Susskind Hamiltonian~\cite{Kogut:1974ag}, which is constructed in the temporal gauge. An important feature of this Hamiltonian is the Gauss law constraint, which has to be imposed on physical states. The Gauss law constraint equation commutes with the Hamiltonian, which means that if the Gauss law is initially imposed, in principle the state remains physical throughout the time evolution. However, unphysical states may be produced due to Trotterization and hardware error in quantum simulation. One method of suppressing the Gauss law violation in time evolution is to modify the Hamiltonian~\cite{Halimeh:2021lnv}. Another is to use a fault-tolerant setup~\cite{Spagnoli:2024mib}. On the other hand, one can completely fix the Hilbert space to just contain the physical states. This was first demonstrated for SU(2) and SU(3) pure gauge theories on plaquette chains by using the irrep basis~\cite{Klco:2019evd,Ciavarella:2021nmj,ARahman:2021ktn,ARahman:2022tkr}. Tessellations only consisting of three-link vertices can significantly simplify the projection onto physical states.
This line of thinking was highlighted by the recent work that proposed the honeycomb lattice for 2+1D and triamond and hyperhoneycomb lattices for 3+1D~\cite{Muller:2023nnk,Kavaki:2024ijd,Illa:2025dou}.
Other approaches of dealing with the Gauss law in the Kogut-Susskind Hamiltonian includes gauge fixing for spatial gauge fields~\cite{Mariani:2024osg,Grabowska:2024emw,Burbano:2024uvn},\footnote{Under the temporal gauge condition spatial gauge redundancy still exists.} large-$N_c$~\cite{Ciavarella:2024fzw,Ciavarella:2025bsg}, $q$-deformed algebras~\cite{Zache:2023dko,Hayata:2023bgh,Hayata:2024fnh} and changing basis such as the loop-string-hadron formulation~\cite{Raychowdhury:2019iki,Kadam:2022ipf,Kadam:2024ifg}. There are also the quantum link model~\cite{Chandrasekharan:1996ih} and the orbifold~\cite{Bergner:2024qjl,Halimeh:2024bth} as alternative formulations. At the moment, no studies claimed that certain lattice gauge theory simulation problems are BQP or BQP-complete. A crucial aspect of such a claim is to estimate the qubit cost to represent all states on the lattice up to a given energy with a given accuracy. This type of estimate was performed for SU(2) pure gauge theories in the irrep basis, which shows polynomial scaling~\cite{Turro:2024pxu}. Convergence of highly excited states with the irrep basis truncation was numerically demonstrated in Ref.~\cite{Ebner:2023ixq}.

One difficulty in resource estimate for lattice gauge theory simulation using the Kogut-Susskind Hamiltonian is the Gauss law constraint. This motivates considering other gauge fixing conditions and Hamiltonians. For example, 1+1D Hamiltonians in the axial gauge have been studied~\cite{Farrell:2022wyt,Farrell:2023fgd}. For the Schwinger model, it can be explicitly shown equivalent to the Kogut-Susskind Hamiltonian if the Gauss law is completely integrated out.\footnote{This means expressing the electric fields completely in terms of the fermion fields, which can be done with aperiodic fixed boundary conditions.} A more recent work considered the Hamiltonian of Quantum Electrodynamics (QED), which is in 3+1D, in the Coulomb gauge and proposed representing the gauge degrees of freedom in the occupation basis in momentum space~\cite{Li:2024ide}.

Here we consider the QED Hamiltonian in the Coulomb gauge again but representing the gauge degrees of freedom in the field basis in position space, which is more efficient. This enables us to estimate the number of qubits needed to represent physical states up to a given energy with a given accuracy, which is as systematic and rigorous as the JLP studies for scalar field theories. We prove that the qubit cost scales polynomially with the lattice size, energy, accuracy, and Hamiltonian parameters in lattice units.
Equations~\eqref{eqn:Amax},~\eqref{eqn:Pimax} and~\eqref{eqn:nA_final} are new results obtained here for bosonic fields coupled with fermions.
In both the continuum and the discretized
Coulomb gauge Hamiltonians, it is manifest that the unphysical longitudinal gauge fields
are decoupled and commute with the corresponding Hamiltonian.
So there is no need to impose any constraint, contrary to the case of the Kogut-Susskind Hamiltonian. For quantum simulation, the field basis and conjugate variable basis at one lattice site can be swapped efficiently by using the quantum Fourier transform algorithm. We further show that the gate cost for real-time quantum simulation scales polynomially with the lattice size, energy, time, accuracy, and Hamiltonian parameters in lattice units. Both single-qubit and two-qubit gate costs here for implementing the gauge field time evolution are polynomially reduced compared with those in the previous work using the occupancy basis in momentum space.
Here we focus on the lattice theory without discussing the ultraviolet (UV) completion of QED. The total resources needed for a continuum extrapolation still scale polynomially as long as the extrapolation does not require an exponential number of data points at finite lattice spacings.

This paper is organized as follows: In Sec.~\ref{sec:H_QED}, we will briefly review the continuum QED Hamiltonian in the Coulomb gauge, and show that it is equivalent to that in the temporal gauge for physical states. In Sec.~\ref{sec:lattice}, a lattice discretization will be
introduced that maintains the decoupling of the longitudinal gauge fields and their commuting with the Hamiltonian. The issue of continuum limit will be discussed.
Maps of both gauge and fermion field degrees of freedom in position space onto qubits will also be given. 
We will prove the polynomial scaling in the qubit cost to represent physical states in Sec.~\ref{sec:qubit}, followed by an estimate of the gate count for quantum simulation of real-time evolution in Sec.~\ref{sec:real-time}, which will also be shown to scale polynomially. Furthermore, we compare the gate costs with those in the previous work in Sec~\ref{sec:comparison} and demonstrate the significant advantage of using the field basis in position space considered here. Finally, we will summarize and give an outlook in Sec.~\ref{sec:conclusions}.

\section{QED Hamiltonian}
\label{sec:H_QED}
In this section, we first briefly review the continuum QED Hamiltonians in the Coulomb and temporal gauge conditions and then show they are equivalent for physical states.

We use the most negative convention for the Minkowski metric $\eta_{\mu\nu}={\rm diag}(1,-1,-1,-1)$. We follow the convention that Greek letters ($\mu,\nu,\cdots$) label Lorentzian indices while Roman letters ($i,j,\cdots$) stand for spatial indices.  Repeated indices are contracted. Contractions of one upper and one lower indices are of Lorentzian type while contractions of two lower indices are of Euclidean type.

\subsection{Hamiltonian in Coulomb Gauge}
\label{sec:H_coulomb}
The Hamiltonian of QED in the Coulomb gauge $\partial_i A_i = 0$ can be obtained from the Dirac quantization of constrained systems. Details can be found in standard references, e.g., Ref.~\cite{Weinberg:1995mt}. The Hamiltonian with one flavor of fermions can be written as
\begin{subequations}
\label{eqn:H_coulomb}
\bea
H &= \int {\rm d}^3x \,\mathcal{H}({\bs x}) \,, \\
\mathcal{H}({\boldsymbol x}) &= \frac{1}{2}\Pi_{\perp i}^2({\boldsymbol x}) + \frac{1}{2}[\varepsilon_{ijk}\partial_j A_k({\boldsymbol x})]^2 -J^i({\boldsymbol x})A_i({\boldsymbol x}) - \frac{1}{2}J^0({\boldsymbol x})A_0({\boldsymbol x}) - \bar{\psi}({\boldsymbol x})(i\gamma^i\partial_i - m)\psi({\boldsymbol x})  \,,
\eea
\end{subequations}
where $\varepsilon_{ijk}$ denotes the Levi-Civita tensor and $A_i$ for $i\in\{1,2,3\}$ are dynamical gauge field variables with the canonically conjugate variables $\Pi_{\perp i}$. Their commutation relations are given by promoting the classical Dirac bracket to the quantum commutator
\begin{subequations}
\label{eqn:commu_A}
\bea
\label{eqn:commu_AA}
[A_i(\boldsymbol x), A_j(\boldsymbol y)] &= 0 \,, \\
[\Pi_{\perp i}(\boldsymbol x), \Pi_{\perp j}(\boldsymbol y)] &= 0 \,,\\
\label{eqn:commu_APi}
[A_i(\boldsymbol x), \Pi_{\perp j}(\boldsymbol y)] &= i \left(\delta_{ij}-\frac{\partial_i\partial_j}{\nabla^2} \right)\delta^{(3)}({\boldsymbol x} - {\boldsymbol y}) = i\delta_{ij} \delta^{(3)}({\boldsymbol x} - {\boldsymbol y}) + i \partial_i\partial_j \frac{1}{4\pi|{\boldsymbol x} - {\boldsymbol y}|} \,,
\eea
\end{subequations}
where derivatives act on $\bs x$. The gauge potential $A_0$ is uniquely fixed by $\nabla^{-2} J_0$ under the boundary conditions that all fields vanish at spatial infinity, which we will use throughout. The Coulomb potential interaction term is then given by
\bea
\label{eqn:J0A0}
-\frac{1}{2} \int d^3x J^0(\boldsymbol x) A_0(\boldsymbol x) = -\frac{1}{2} \int d^3x J^0(\boldsymbol x) \frac{1}{\nabla^2} J_0(\boldsymbol x) = \frac{1}{2} \int d^3x \int d^3y \frac{J^0(\boldsymbol x) J_0(\boldsymbol y)}{4\pi|{\boldsymbol x}-{\boldsymbol y}|} \,.
\eea
The fermion electromagnetic current density is given by
\bea
J^\mu(\boldsymbol x) = g \bar{\psi}(\boldsymbol x) \gamma^\mu \psi(\boldsymbol x) \,,
\eea
where $g$ denotes the coupling and $\psi$ is the fermion field with mass $m$ and $\bar{\psi} = \psi^\dagger \gamma^0$. They obey the standard anticommutation relations
\begin{subequations}
\label{eqn:commu_psi}
\bea
\{\psi_\alpha(\boldsymbol x), \psi_\beta(\boldsymbol y)\} &= 0\,, \\
\{\psi^\dagger_\alpha(\boldsymbol x), \psi^\dagger_\beta(\boldsymbol y)\} &= 0\,, \\
\{\psi_\alpha(\boldsymbol x), \psi^\dagger_\beta(\boldsymbol y)\} &= \delta_{\alpha\beta}\delta^{(3)}(\boldsymbol x-\boldsymbol y) \,.
\eea
\end{subequations}
The Dirac fermion in 3+1D has four components labeled by $\alpha$ and $\beta$ here.

The Dirac bracket accounts for two second-class constraints 
\begin{subequations}
\label{eqn:constraints_coulomb}
\bea
\partial_i A_i &= 0 \,,\\
\partial_i \Pi_{\perp i}  &= 0\,,
\eea
\end{subequations}
which should be thought of as constraints on the dynamical variables rather than those imposed on physical states as in the case of the temporal gauge. The former constraint $\partial_i A_i = 0$ is the Coulomb gauge condition. The latter constraint $\partial_i \Pi_{\perp i} = 0$ is obtained from the Euler Lagrangian equation for $A_0$ and using the requirement that $F_{00}=0$ is consistent with the time evolution. Because of these two constraints, only two gauge field variables are independent. The standard canonical quantization expresses the Hamiltonian in terms of these two independent gauge field variables and their corresponding canonically conjugate variables. They follow the standard commutation relations given by promoting the Poisson bracket to the quantum commutator. The Dirac quantization procedure treats all three spatial gauge fields as dynamical variables and expresses the Hamiltonian in terms of them and their corresponding conjugates, which follow the commutation relations given by the Dirac bracket. The commutators~\eqref{eqn:commu_A} are consistent with these two constraints and we no longer need to consider the constraints in Eq.~\eqref{eqn:constraints_coulomb} when using the Coulomb gauge Hamiltonian, which will be explained in detail in Sec.~\ref{H_equivalence}. We show the quantized Hamiltonian obtained by just using two independent variables in Appendix~\ref{app} and discuss some challenges for numerical simulation, which to our knowledge was not explained anywhere before.

\subsection{Hamiltonian in Temporal Gauge}
\label{sec:H_temporal}
The temporal gauge condition $A_0=0$ is used e.g., in the Kogut-Susskind formulation of lattice Hamiltonians for gauge theories~\cite{Kogut:1974ag}. The QED Hamiltonian density is given by
\bea
\label{eqn:H_temporal}
\mathcal{H}(\boldsymbol x) = \frac{1}{2} \Pi_i^2({\boldsymbol x}) + \frac{1}{2}[\varepsilon_{ijk}\partial_j A_k({\boldsymbol x})]^2 -J^i({\boldsymbol x})A_i({\boldsymbol x}) - \bar{\psi}({\boldsymbol x})(i\gamma^i\partial_i - m)\psi({\boldsymbol x})  \,.
\eea
The gauge field variables and their canonically conjugate variables follow the commutation relations
\begin{subequations}
\bea
[A_i(\boldsymbol x), A_j(\boldsymbol y)] &= 0 \,, \\
[\Pi_{i}(\boldsymbol x), \Pi_{j}(\boldsymbol y)] &= 0 \,,\\
\label{eqn:temporal_commu}
[A_i(\boldsymbol x), \Pi_{j}(\boldsymbol y)] &= i \delta_{ij} \delta^{(3)}({\boldsymbol x} - {\boldsymbol y}) \,.
\eea
\end{subequations}
The fermion fields follow the same anticommutation relations as in Eq.~\eqref{eqn:commu_psi}.

The Gauss law constraint can be written as
\bea
\label{eqn:gauss}
\partial_i \Pi_i \approx J^0 \,.
\eea
The physical meaning of $\Pi_i$ is the electric field. The approximation sign here means that it is imposed on physical states rather than an operator identity. More specifically, we have
\bea
(\partial_i \Pi_i - J^0) |\Psi^{\rm Phys}\rangle = 0\,.
\eea
A direct way of seeing this, i.e., Eq.~\eqref{eqn:gauss} is not an operator identity, is to note that it is not consistent with the commutation relation~\eqref{eqn:temporal_commu}, to which applying $\partial/\partial y_j$ gives
\bea
\label{eqn:gauss_inconsistency}
[A_i(\boldsymbol x), \partial_j\Pi_{j}(\boldsymbol y)] = i \frac{\partial}{\partial y_i} \delta^{(3)}({\boldsymbol x} - {\boldsymbol y}) \neq 0\,.
\eea
However, $A_i$, $\psi_\alpha$, and $\psi^\dagger_\alpha$ are treated as independent variables in the canonical quantization, so we have
\bea
[A_i(\boldsymbol x), J^0(\boldsymbol y)] = [A_i(\boldsymbol x), g\bar{\psi}(\boldsymbol y)\gamma^0\psi(\boldsymbol y)] = 0 \,,
\eea
which would be inconsistent with Eq.~\eqref{eqn:gauss_inconsistency}, if Eq.~\eqref{eqn:gauss} were treated as an operator identity. On the other hand, everything is consistent when the constraint is only imposed on physical states. In particular, we have
\bea
[A_i(\boldsymbol x), \partial_j\Pi_{j}(\boldsymbol y)] |\Psi^{\rm Phys}\rangle = [\partial_j\Pi_{j}(\boldsymbol y) - J^0({\boldsymbol y}) ]A_i(\boldsymbol x) |\Psi^{\rm Phys}\rangle
\neq 0 \,,
\eea
which just states that $A_i(\boldsymbol x) |\Psi^{\rm Phys}\rangle$ is not a physical state.

\subsection{Equivalence for Physical States}
\label{H_equivalence}
In this subsection, we show that the Hamiltonians in the Coulomb and temporal gauge conditions are equivalent when acting on physical states. We differentiate gauge field variables in these two gauge conditions by using the superscripts $(c)$ and $(t)$. Our starting point is the Hamiltonian in the temporal gauge and we decompose the electric field into a transverse and a longitudinal part
\bea
\Pi^{(t)}_i(\boldsymbol x) = \Pi^{(t)}_{\perp i}(\boldsymbol x) + \Pi^{(t)}_{\parallelsum\,i}(\boldsymbol x)\,.
\eea
These two parts can be formally written as
\begin{subequations}
\bea
\Pi^{(t)}_{\perp i}(\boldsymbol x) &= \left(\delta_{ij} - \frac{\partial_i\partial_j}{\nabla^2}\right) \Pi^{(t)}_j(\boldsymbol x)  \,,\\
\Pi^{(t)}_{\parallelsum\,i}(\boldsymbol x) &= \frac{\partial_i\partial_j}{\nabla^2} \Pi^{(t)}_j(\boldsymbol x) \,,
\eea
\end{subequations}
which are mathematically defined in momentum space. Then the Gauss law constraint can be written as
\begin{subequations}
\label{eqn:gauss_TL}
\bea
\partial_i \Pi^{(t)}_{\perp i} & = 0 \,,\\
\partial_i \Pi^{(t)}_{\parallelsum\,i} & \approx J^0 \,,
\eea
\end{subequations}
where the approximation sign again means that the second constraint is imposed on physical states. When the electric energy operator, i.e., the first term in Eq.~\eqref{eqn:H_temporal} integrated over the whole space, acts on a physical state, we have
\bea
\label{eqn:E2_phys}
\frac{1}{2}\!\int\! {\rm d}^3x[\Pi_i^{(t)}({\boldsymbol x})]^2 \approx \frac{1}{2}\!\int\! {\rm d}^3x\!\left[\Pi_{\perp i}^{(t)}({\boldsymbol x}) + \frac{\partial_i}{\nabla^2} J^0({\boldsymbol x}) \right]^2 = \frac{1}{2}\!\int\! {\rm d}^3x \!\left\{[\Pi_{\perp i}^{(t)}({\boldsymbol x})]^2 - J^0({\boldsymbol x})\frac{1}{\nabla^2}J^0({\boldsymbol x}) \right\} \,,
\eea
where we have used Eq.~\eqref{eqn:gauss_TL} and the boundary conditions that both gauge and fermion fields vanish at spatial infinity. We recognize the last term in Eq.~\eqref{eqn:E2_phys} as exactly the Coulomb interaction term~\eqref{eqn:J0A0} in the Coulomb gauge Hamiltonian.

What remains to show is how the Dirac bracket commutation relation~\eqref{eqn:commu_APi} realizes that $\Pi^{(c)}_\perp$ in the electric energy in the Coulomb gauge Hamiltonian is transverse. If we introduce a conjugate variable $\Pi_i^{(c)}$ that satisfies
\begin{subequations}
\label{eqn:commu_A_Pinew}
\bea
[\Pi_i^{(c)}(\boldsymbol x), \Pi_j^{(c)}(\boldsymbol y)] &=0 \,,\\
[A_i^{(c)}(\boldsymbol x), \Pi_j^{(c)}(\boldsymbol y)] &= i\delta_{ij}\delta^{(3)}(\boldsymbol x - \boldsymbol y) \,,
\eea
\end{subequations}
we will see that the original conjugate variable in the Coulomb gauge Hamiltonian, $\Pi_{\perp i}^{(c)}$ that follows the commutation relations~\eqref{eqn:commu_A}, is given by
\bea
\Pi_{\perp i}^{(c)}(\boldsymbol x) = \left( \delta_{ij} - \frac{\partial_i \partial_j} {\nabla^2} \right) \Pi_j^{(c)}(\boldsymbol x) \,.
\eea
So $\Pi_{\perp i}^{(c)}$ is transverse.
In terms of the new conjugate variable $\Pi_{i}^{(c)}$, the electric energy in the Coulomb gauge Hamiltonian becomes
\bea
\label{eqn:H_E_coulomb}
\frac{1}{2} \int {\rm d}^3x[\Pi_{\perp i}^{(c)}({\boldsymbol x})]^2 = \frac{1}{2} \int {\rm d}^3x \, \Pi_i^{(c)}(\boldsymbol x)  \left( \delta_{ij} - \frac{\partial_i \partial_j} {\nabla^2} \right) \Pi_j^{(c)}(\boldsymbol x) \,. 
\eea
We see that only the transverse part of $\Pi_i^{(c)}$ contributes to the electric energy. Equation~\eqref{eqn:H_E_coulomb} in the Coulomb gauge is equivalent to the second-to-last term in Eq.~\eqref{eqn:E2_phys} for the temporal gauge case since by definition $\partial_i\Pi_{\perp i}^{(t)} = 0$ as in Eq.~\eqref{eqn:gauss_TL}.

In a nutshell, we showed that the Hamiltonian in the Coulomb gauge is equivalent to that in the temporal gauge when acting on physical states. In the Coulomb gauge Hamiltonian, the magnetic energy term is given by
\bea
\label{eqn:H_AA term}
\frac{1}{2}\int {\rm d}^3x [\varepsilon_{ijk}\partial_j A_k(\boldsymbol x)]^2 = - \frac{1}{2}\int {\rm d}^3x A_i(\boldsymbol x) (\delta_{ij}\nabla^2 - \partial_i\partial_j) A_j(\boldsymbol x) \,,
\eea
to which only the transverse component of the gauge field contributes. Together with Eq.~\eqref{eqn:H_E_coulomb}, we conclude that only the transverse gauge fields can propagate and the constraints in Eq.~\eqref{eqn:constraints_coulomb} are accounted for by the Dirac bracket commutation relations and the Hamiltonian, so one does not need to impose any constraint, contrary to the temporal gauge case. In the current way of writing, $J^iA_i$ may contain contributions from the longitudinal gauge fields, but two $J^i$s are not coupled via the propagation of the longitudinal modes. So these contributions do not lead to nontrivial dynamics. To explicitly remove these irrelevant phases in the Hamiltonian evolution, we can replace $A_i$ with $(\delta_{ij}-\partial_i\partial_j/\nabla^2)A_j$ in the term $J^iA_i$
or impose $\partial_i A_i(\boldsymbol{x})=0, \forall {\bs x}$ as an initial condition, since the longitudinal gauge fields are dynamically decoupled $[\partial_i A_i(\boldsymbol{x}), H] = 0$.

\section{Lattice Formulation in Field Basis and Map onto Qubits}
\label{sec:lattice}
We consider a 3D spatial cubic lattice of size $L$ along each direction and volume $V=L^3$ with lattice spacing $a$. In total, the lattice has $\hat{V}=\hat{L}^3$ sites where $\hat{L}=L/a$. Spatial points are labeled by
\bea
\boldsymbol x = \hat{\boldsymbol x}a = (\hat{x}_1 a, \hat{x}_2 a, \hat{x}_3 a)\,,\ \hat{x}_i\in \{1,2,3,\cdots,\hat{L} \} \,.
\eea
We will use the Dirichlet boundary conditions for the fields in the following, so the lattice is not periodic. As will be explained shortly in Sec.~\ref{sec:lattice_qed}, the eigenvalues and eigenstates of the discrete Laplacian operator can be written in terms of discrete momenta specified by
\bea
\label{eqn:discrete_momentum_Dirichlet}
\boldsymbol p = \frac{\hat{\boldsymbol p}}{a} = \frac{\pi}{(\hat{L}+1)a}(k_1,k_2,k_3) \,,\ k_i\in \{1,2,3,\cdots,\hat{L} \} \,.
\eea

\subsection{Lattice QED Hamiltonian in Coulomb Gauge}
\label{sec:lattice_qed}
To properly discretize the Coulomb gauge QED Hamiltonian, we need to define the meaning of $1/\nabla^2$ on the lattice, i.e., construct the Green's function for the discrete Laplacian operator. To this end, we introduce some notations for discrete differences
\begin{subequations}
\bea
\partial_{\hat{x}_i}^{(L+)} f(\hat{\boldsymbol x}) &\equiv  f(\hat{\boldsymbol x}+\hat{i}) - f(\hat{\boldsymbol x}) \,,\\
\partial_{\hat{x}_i}^{(L-)} f(\hat{\boldsymbol x}) &\equiv  f(\hat{\boldsymbol x}) - f(\hat{\boldsymbol x}-\hat{i}) \,,
\eea
\end{subequations}
where the superscript $L$ indicates ``lattice'' and $\hat{i}$ denotes a unit vector along the $i$ axis.
Under the Dirichlet boundary conditions for functions $f$ and $g$, i.e., $f(\hat{\boldsymbol x})=g(\hat{\boldsymbol x}) = 0$ if $\hat{x}_i = 0$ or $L+1$ for any $i$, one can easily prove the discrete version of ``integration by parts''
\bea
\sum_{\hat{\boldsymbol{x}}} f(\hat{\boldsymbol x}) \partial_{\hat{x}_i}^{(L+)} g(\hat{\boldsymbol x}) = - \sum_{\hat{\boldsymbol{x}}} \big[\partial_{\hat{x}_i}^{(L-)} f(\hat{\boldsymbol x}) \big] g(\hat{\boldsymbol x}) \,,
\eea
where the square bracket on the right-hand side emphasizes that the operator $\partial_{\hat{x}_i}^{(L-)}$ only acts on $f(\hat{\boldsymbol x})$.

We now define the discrete Laplacian operator as
\bea
\label{eqn:discrete_Laplacian}
\big[\nabla_{\hat{\boldsymbol{x}}}^{(L)}\big]^2 \equiv \sum_i \partial_{\hat{x}_i}^{(L+)} \partial_{\hat{x}_i}^{(L-)} = \sum_i \partial_{\hat{x}_i}^{(L-)} \partial_{\hat{x}_i}^{(L+)} \,,
\eea
which acts on a function $f$ at site $\hat{\boldsymbol{x}}$ as
\bea
\big[\nabla_{\hat{\boldsymbol{x}}}^{(L)}\big]^2 f(\hat{\boldsymbol{x}}) = \sum_i \big[ f(\hat{\boldsymbol{x}}+\hat{i}) + f(\hat{\boldsymbol{x}}-\hat{i}) - 2 f(\hat{\boldsymbol{x}}) \big]
\eea
Using the method discussed in Ref.~\cite{chung2000discrete}, we find the Green's function for the discrete Laplacian operator $\big[\nabla_{\hat{\boldsymbol{x}}}^{(L)}\big]^2$ is
\bea
\label{eqn:Green for discrete Laplacian}
G(\hat{\boldsymbol{x}}, \hat{\boldsymbol{y}}) = \sum_{\hat{\boldsymbol{p}}} \frac{\phi_{\hat{p}_1}(\hat{x}_1)\phi_{\hat{p}_1}(\hat{y}_1) \phi_{\hat{p}_2}(\hat{x}_2)\phi_{\hat{p}_2}(\hat{y}_2) \phi_{\hat{p}_3}(\hat{x}_3)\phi_{\hat{p}_3}(\hat{y}_3) }{\lambda_{\hat{p}_1}+\lambda_{\hat{p}_2}+\lambda_{\hat{p}_3}} \,,
\eea
where $\sum_{\hat{\boldsymbol{p}}}$ means $\sum_{k_1,k_2,k_3}$ with $k_i\in \{1,2,3,\cdots,\hat{L} \}$ as in Eq.~\eqref{eqn:discrete_momentum_Dirichlet}. The eigenvalues and orthonormal eigenvectors of the operator $\partial_{\hat{x}_i}^{(L+)} \partial_{\hat{x}_i}^{(L-)}$ are 
\begin{subequations}
\bea
\lambda_{\hat{p}_i} &= 2\cos\hat{p}_i - 2 = 2\cos\frac{k_i\pi}{\hat{L}+1} - 2 \,,\\
\phi_{\hat{p}_i}(\hat{x}_i) &= \sqrt{\frac{2}{\hat{L}+1}}\sin(\hat{p}_i \hat{x}_i) = \sqrt{\frac{2}{\hat{L}+1}}\sin\frac{k_i \hat{x}_i \pi}{\hat{L}+1} \,, \label{eqn:phi_k(x)}
\eea
\end{subequations}
where there is no summation over $i$. From the definition, one can easily see that the Green's function is symmetric in $\hat{\boldsymbol{x}}$ and $\hat{\boldsymbol{y}}$, i.e., $G(\hat{\boldsymbol{x}}, \hat{\boldsymbol{y}}) = G(\hat{\boldsymbol{y}}, \hat{\boldsymbol{x}})$. Furthermore, it satisfies
\bea
\label{eqn:green function delta}
\big[\nabla_{\hat{\boldsymbol{x}}}^{(L)}\big]^2 G(\hat{\boldsymbol{x}}, \hat{\boldsymbol{y}}) = \big[\nabla_{\hat{\boldsymbol{y}}}^{(L)}\big]^2 G(\hat{\boldsymbol{x}}, \hat{\boldsymbol{y}}) = \delta_{\hat{\boldsymbol{x}} \hat{\boldsymbol{y}}} \,,
\eea
where $\delta_{\hat{\boldsymbol x}\hat{\boldsymbol y}}$ denotes a Kronecker delta function defined by
\bea
\delta_{\hat{\boldsymbol x}\hat{\boldsymbol y}} = 
\begin{cases}
    1\,, & \text{if } \hat{\boldsymbol x}=\hat{\boldsymbol y} \\
    0\,, & \text{if } \hat{\boldsymbol x}\neq \hat{\boldsymbol y} 
\end{cases} \,.
\eea

With these preparations, we can now write down a lattice QED Hamiltonian in the Coulomb gauge on the spatial lattice by discretizing Eqs.~\eqref{eqn:H_coulomb},~\eqref{eqn:J0A0},~\eqref{eqn:H_E_coulomb}, and~\eqref{eqn:H_AA term},
\begin{subequations}
\label{eqn:latticeH}
\bea
\hat{H}& = aH = \hat{H}_\Pi + \hat{H}_A + \hat{H}_C + \hat{H}_f \,, \\
\hat{H}_\Pi &= \frac{1}{2} \sum_{\hat{\boldsymbol x},i}  \hat{\Pi}_{i}^2(\hat{\boldsymbol x}) + \frac{1}{2} \sum_{\hat{\boldsymbol x},\hat{\boldsymbol y},i,j}  \hat{\Pi}_{i}(\hat{\boldsymbol x}) \hat{\Pi}_{j}(\hat{\boldsymbol y}) \partial_{\hat{x}_i}^{(L+)} \partial_{\hat{y}_j}^{(L+)} G(\hat{\boldsymbol{x}}, \hat{\boldsymbol{y}})  \,, \label{eqn:Hpi} \\
%\hat{H}_\Pi &= \frac{1}{2} \sum_{\hat{\boldsymbol x},i}  \hat{\Pi}_{i}^2(\hat{\boldsymbol x}) + \frac{1}{2} \sum_{\hat{\boldsymbol x},\hat{\boldsymbol y},i,j} \frac{\hat{\Pi}_{i}(\hat{\boldsymbol x}) [\hat{\Pi}_j(\hat{\boldsymbol y}+\hat{i}+\hat{j}) -\hat{\Pi}_j(\hat{\boldsymbol y}+\hat{i}) -\hat{\Pi}_j(\hat{\boldsymbol y}+\hat{j}) + \hat{\Pi}_j(\hat{\boldsymbol y})] }{4\pi|\hat{\boldsymbol x}-\hat{\boldsymbol y}|} \,, \label{eqn:Hpi}\\
\hat{H}_A & = - \frac{1}{2}  \sum_{\hat{\boldsymbol x},i}\hat{A}_i(\hat{\boldsymbol x}) \big[\nabla_{\hat{\boldsymbol{x}}}^{(L)}\big]^2 \hat{A}_i(\hat{\boldsymbol x}) + \frac{1}{2}\sum_{\hat{\boldsymbol x}, i,j} \hat{A}_i(\hat{\boldsymbol x}) \partial_{\hat{x}_i}^{(L+)} \partial_{\hat{x}_j}^{(L-)} \hat{A}_j(\hat{\boldsymbol x}) \,, \label{eqn:HA}\\
%\hat{H}_A & = - \frac{1}{2}  \sum_{\hat{\boldsymbol x},i,j}\hat{A}_j(\hat{\boldsymbol x}) \Big[ \hat{A}_j(\hat{\boldsymbol x}+\hat{i}) -2\hat{A}_j(\hat{\boldsymbol x}) + \hat{A}_j(\hat{\boldsymbol x}-\hat{i}) \Big] \nn\\
%& \quad + \frac{1}{2}\sum_{\hat{\boldsymbol x}, i,j} \hat{A}_i(\hat{\boldsymbol x}) \Big[ \hat{A}_j(\hat{\boldsymbol x}+\hat{i}+\hat{j}) -\hat{A}_j(\hat{\boldsymbol x}+\hat{i}) -\hat{A}_j(\hat{\boldsymbol x}+\hat{j}) + \hat{A}_j(\hat{\boldsymbol x}) \Big] \,, \label{eqn:HA} \\
\hat{H}_I &= -\sum_{\hat{\boldsymbol x},i} g\bar{\hat{\psi}}(\hat{\boldsymbol x}) \gamma^i \hat{\psi}(\hat{\boldsymbol x}) \hat{A}_i(\hat{\boldsymbol x}) \,
\label{eqn:HI} \\
\hat{H}_C &= -\frac{g^2}{2} \sum_{\hat{\boldsymbol x},\hat{\boldsymbol y}} G(\hat{\boldsymbol x},\hat{\boldsymbol y}) \bar{\hat{\psi}}(\hat{\boldsymbol x}) \gamma^0 \hat{\psi}(\hat{\boldsymbol x}) \bar{\hat{\psi}}(\hat{\boldsymbol y}) \gamma^0 \hat{\psi}(\hat{\boldsymbol y}) \,, \label{eqn:HC} \\
%\hat{H}_C &= \frac{g^2}{2} \sum_{\hat{\boldsymbol x}} \sum_{\hat{\boldsymbol y}\neq \hat{\boldsymbol x}} \frac{ \bar{\hat{\psi}}(\hat{\boldsymbol x}) \gamma^0 \hat{\psi}(\hat{\boldsymbol x}) \bar{\hat{\psi}}(\hat{\boldsymbol y}) \gamma^0 \hat{\psi}(\hat{\boldsymbol y}) }{4\pi|\hat{\boldsymbol x}-\hat{\boldsymbol y}|} \,, \label{eqn:HC} \\
\hat{H}_f &= - \frac{i}{2} \sum_{\hat{\boldsymbol x},j} \bar{\hat{\psi}}(\hat{\boldsymbol x}) \gamma^j \Big[ \hat{\psi}(\hat{\boldsymbol x}+\hat{j}) - \hat{\psi}(\hat{\boldsymbol x}-\hat{j}) \Big] + \hat{m} \sum_{\hat{\boldsymbol x}} \bar{\hat{\psi}}(\hat{\boldsymbol x}) \hat{\psi}(\hat{\boldsymbol x}) -\frac{\hat{r}}{2} \sum_{\hat{\boldsymbol x},i} \bar{\hat{\psi}}(\hat{\boldsymbol x}) \big[\nabla_{\hat{\boldsymbol{x}}}^{(L)}\big]^2 \hat{\psi}(\hat{\boldsymbol x}) \,,
%& \quad -\frac{\hat{r}}{2} \sum_{\hat{\boldsymbol x},i} \bar{\hat{\psi}}(\hat{\boldsymbol x}) \Big[ \hat{\psi}(\hat{\boldsymbol x}+\hat{i}) -2\hat{\psi}(\hat{\boldsymbol x}) + \hat{\psi}(\hat{\boldsymbol x}-\hat{i}) \Big] \label{eqn:Hf} \,,
\eea
\end{subequations}
where all variables are scaled by proper powers of $a$ and thus made unitless, i.e., $\hat{A}_i = aA_i$, $\hat{\Pi}_{i} = a^2\Pi_{i}$, $\hat{\psi} = a^{3/2}\psi$, and $\hat{m}=am$. We have used Eq.~\eqref{eqn:H_E_coulomb} and the Green's function of the discrete Laplacian operator in Eq.~\eqref{eqn:Green for discrete Laplacian} when writing down the electric energy term $\hat{H}_\Pi$. In the fermion Hamiltonian $H_f$, we added a Wilson term with the unitless positive coefficient $\hat{r}>0$ to avoid the fermion doubling problem~\cite{Jordan:2014tma}.

The gauge field commutation relations in Eqs.~\eqref{eqn:commu_AA} and~\eqref{eqn:commu_A_Pinew} are modified to be
\begin{subequations}
\label{eqn:commu_A_lattice}
\bea
[\hat{A}_i(\hat{\boldsymbol x}), \hat{A}_j(\hat{\boldsymbol y})] &= 0 \,, \\
[\hat{\Pi}_{i}(\hat{\boldsymbol x}), \hat{\Pi}_{j}(\hat{\boldsymbol y})] &= 0 \,,\\
\label{eqn:commu_APi_lattice}
[\hat{A}_i(\hat{\boldsymbol x}), \hat{\Pi}_{j}(\hat{\boldsymbol y})] &= i \delta_{ij} \delta_{\hat{\boldsymbol x}\hat{\boldsymbol y}} \,.
\eea
\end{subequations}
The fermion field anticommutation relations in Eq.~\eqref{eqn:commu_psi} are modified as
\begin{subequations}
\label{eqn:commu_psi_lattice}
\bea
\{\hat{\psi}_\alpha(\hat{\boldsymbol x}), \hat{\psi}_\beta(\hat{\boldsymbol y})\} &= 0\,, \\
\{\hat{\psi}^\dagger_\alpha(\hat{\boldsymbol x}), \hat{\psi}^\dagger_\beta(\hat{\boldsymbol y})\} &= 0\,, \\
\{\hat{\psi}_\alpha(\hat{\boldsymbol x}), \hat{\psi}^\dagger_\beta(\hat{\boldsymbol y})\} &= \delta_{\alpha\beta}\delta_{\hat{\boldsymbol x} \hat{\boldsymbol y}} \,.
\eea
\end{subequations}

\subsection{Gauge Invariance and Continuum Limit}
\label{sec:lattice gauge invariance}
We note that in the lattice QED Hamiltonian~\eqref{eqn:latticeH}, the gauge field degrees of freedom appear as $\hat{A}_i$, which are not gauge covariant or invariant, contrary to the spatial Wilson lines and loops in the Kogut-Susskind Hamiltonian. Yet we need to make sure that gauge invariance in the original continuum theory is restored.

The physical meaning of gauge invariance is that only the physical degrees of freedom contribute to observables in the end. For the continuum QED Hamiltonian in the Coulomb gauge, gauge invariance means that the longitudinal component of the gauge field $A_{\parallelsum\,i}(\boldsymbol x) = \sum_j \frac{\partial_i\partial_j}{\nabla^2}A_j(\boldsymbol x)$ is decoupled and commutes with the Hamiltonian. Mathematically it means the $A_i(\boldsymbol x) (\delta_{ij}\nabla^2 - \partial_i\partial_j) A_j(\boldsymbol x)$ term in the Hamiltonian vanishes for longitudinal gauge fields and
\bea
\label{eqn:AL=0 continuum}
\Big[\sum_i \partial_i A_i(\boldsymbol x), H \Big] = 0\,,\  \forall \boldsymbol x \,,
\eea
which can be seen from Eqs.~\eqref{eqn:commu_A_Pinew} and~\eqref{eqn:H_E_coulomb}.

For the discretized Hamiltonian in Eq.~\eqref{eqn:latticeH}, we find
\bea
\label{eqn:AL=0 discrete}
\Big[\sum_i \partial_{\hat{x}_i}^{(L-)} \hat{A}_i(\hat{\boldsymbol x}), \hat{H} \Big] = 0\,,\ \forall \hat{\boldsymbol x} \,,
\eea
which can be shown by using Eqs.~\eqref{eqn:commu_A_lattice},~\eqref{eqn:discrete_Laplacian} and~\eqref{eqn:green function delta}. Equation~\eqref{eqn:AL=0 discrete} means that the lattice version of the longitudinal gauge field
\bea
\hat{A}_{\parallelsum\,i}(\hat{\boldsymbol x}) = \sum_{\hat{\boldsymbol y},j} \partial_{\hat{x}_i}^{(L+)} G(\hat{\boldsymbol x},\hat{\boldsymbol y})  \partial_{\hat{y}_j}^{(L-)} A_j(\boldsymbol y) \,,
\eea
commutes with the lattice Hamiltonian and remains decoupled in the time evolution since $\hat{H}_A$ vanishes for $\hat{A}_{\parallelsum\,i}(\hat{\boldsymbol x})$. 
If the initial gauge field is transverse in the lattice sense, i.e., $\sum_i \partial_{\hat{x}_i}^{(L-)} \hat{A}_i(\hat{\boldsymbol x})=0$ for all $\hat{\boldsymbol x}$, it will remain transverse throughout the time evolution.  
Equation~\eqref{eqn:AL=0 discrete} is equivalent to Eq.~\eqref{eqn:AL=0 continuum}, up to power corrections in the lattice spacing $a$ that vanish in the limit $a\to0$.\footnote{
This is very similar to the case of the Kogut-Susskind Hamiltonian, which is obtained by discretizing the temporal gauge Hamiltonian. The Gauss law in the continuum limit is 
\begin{align}
\label{eqn:gauss_nonabelian}
    \sum_i D_iE_i(\boldsymbol{x}) = 0\,,
\end{align}
where $D_i$ is the covariant derivative for a non-Abelian gauge theory and $E_i$ denotes the electric field. On the lattice, what is implemented as the lattice version of the Gauss law is
\begin{align}
\label{eqn:lattice_gauss_nonabelian}
    \sum_i \big[ E_i(\hat{\boldsymbol x}+\hat{i}) - E_i(\hat{\boldsymbol x}) \big] =0 \,,
\end{align}
where it is understood that the electric fields at different positions are properly color-transported so one can subtract them, e.g., $E_i(\hat{\boldsymbol x}+\hat{i}) \to U(\hat{\boldsymbol x},\hat{\boldsymbol x}+\hat{i}) E_i(\hat{\boldsymbol x}+\hat{i}) U(\hat{\boldsymbol x}+\hat{i},\hat{\boldsymbol x})$ with $U(\hat{\boldsymbol x},\hat{\boldsymbol y})$ denoting a fundamental spatial Wilson line pointing from $\hat{\boldsymbol y}$ to $\hat{\boldsymbol x}$. Equation~\eqref{eqn:lattice_gauss_nonabelian} is equivalent to Eq.~\eqref{eqn:gauss_nonabelian}, up to power corrections in $a$. In other words, only in the continuum limit, does gauge invariance of the temporal gauge Hamiltonian is restored.}
Therefore, we expect gauge invariance to be restored in the continuum limit.

It is known that perturbative QED calculations give a running coupling $\alpha_{\rm EM}$ that grows with the energy scale and diverges at some high energy scale, known as the Landau pole~\cite{bohr1955niels}. Nonperturbatively, it is speculated that the continuum limit of lattice QED is trivial as the electric charge becomes completely screened~\cite{Gockeler:1997dn,Kim:2001am,Gies:2004hy}. So theoretically the continuum limit of lattice QED may become ill defined. However, many studies have performed lattice QED$+$QCD calculations~\cite{Duncan:1996sq,Blum:2007cy,BMW:2013mpk,Horsley:2015eaa,Giusti:2017dwk,MILC:2018ddw,RBC:2018dos,Kordov:2019oer,DiCarlo:2019thl,Blum:2019ugy,NPLQCD:2020ozd}, from early quenched calculations~\cite{Duncan:1996sq,Blum:2007cy,BMW:2013mpk,Giusti:2017dwk,DiCarlo:2019thl} to later ones using fully dynamical lattice QED~\cite{Horsley:2015eaa,MILC:2018ddw,RBC:2018dos,Kordov:2019oer,Blum:2019ugy,NPLQCD:2020ozd}. The UV nature of lattice QED probably does not affect these numerical studies, because the lattice spacings used correspond to typical QCD energy scales, which are much smaller than the Landau pole. In other words, in the continuum extrapolation of these studies, the running of $\alpha_{\rm EM}$ is very slow and probably not sensitive to the existence of a Landau pole, or the UV scale at which lattice QED becomes ill defined. It is shown that the extracted lattice QED results at the typical QCD energy scales can be effectively described by a low-energy effective theory known as the nonrelativistic QED~\cite{NPLQCD:2020ozd}. All of these numerical lattice QED calculations used noncompact gauge fields, as done here in Eq.~\eqref{eqn:latticeH}. It is known that the lattice QED with compact gauge fields has a confinement-deconfinement phase transition~\cite{Polyakov:1975rs}, but the continuum QED is not confined.

\subsection{Field Basis for Gauge Bosons and Map onto Qubits}
We express the gauge degrees of freedom in terms of the local field basis $|\tilde{A}_i(\hat{\boldsymbol x})\rangle$, which is defined by
\bea
\hat{A}_i(\hat{\boldsymbol x}) |\tilde{A}_j(\hat{\boldsymbol y})\rangle = \delta_{ij}\delta_{\hat{\boldsymbol x}\hat{\boldsymbol y}}\tilde{A}_j(\hat{\boldsymbol y}) |\tilde{A}_j(\hat{\boldsymbol y})\rangle \,,
\eea
i.e., when the field operator acting at the same site along the same spatial direction, it returns the field value there, which can take continuous values from $-\infty$ to $\infty$ due to its bosonic nature. A basis for the gauge field Hilbert space on the lattice can be written as
\bea
\bigotimes_{\hat{\boldsymbol x},i} |\tilde{A}_i(\hat{\boldsymbol x})\rangle \,.
\eea
In practical numerical simulations, we must truncate and digitize the gauge field values, which take values in the range $[-\tilde{A}_{\rm max}, \tilde{A}_{\rm max}]$ at the interval $\delta\!\tilde{A}$. 
The local gauge Hilbert space $\{|\tilde{A}_i(\hat{\bs x})\rangle \}$ for one particular $i$ at a given site consists of $2\tilde{A}_{\rm max}/\delta\!\tilde{A}+1$ discrete levels for field values. We can map these levels onto qubits. The number of qubits needed is  
\bea
\label{eqn:nA}
n_A = \lceil \log_2(2\tilde{A}_{\rm max}/\delta\!\tilde{A}+1) \rceil \,.
\eea
One concrete map is as follows, $|-\tilde{A}_{\rm max}\rangle \to |000\cdots 00\rangle$, $|-\tilde{A}_{\rm max}+\delta \!\tilde{A}\rangle \to |000\cdots 01\rangle$, $|-\tilde{A}_{\rm max}+2\delta\!\tilde{A}\rangle \to |000\cdots 10\rangle$, and so on. Some qubit states may not correspond to any physical states, since $\log_2(2\tilde{A}_{\rm max}/\delta\!\tilde{A}+1)$ may not be an integer. The total number of qubits needed for describing gauge degrees of freedom on the whole lattice is $3n_A \hat{V}$.

In Sec.~\ref{sec:qubit}, we will prove bounds for both $\tilde{A}_{\rm max}$ and $\delta\!\tilde{A}$, which are new results and together give a bound on $n_A$.

\subsection{Field Basis for Fermions and Map onto Qubits}
Following Ref.~\cite{Jordan:2014tma}, we specify the fermion field basis by using the following $4\hat{V}$ commuting observables
\bea
\label{eqn:S_x}
S_x = \{ \hat{\psi}^\dagger_\alpha(\hat{\boldsymbol x})\hat{\psi}_\alpha(\hat{\boldsymbol x}) \ |\ \hat{\boldsymbol x},\, \alpha=1,2,3,4 \} \,,
\eea
where the repeated index $\alpha$ is not summed. Because of the anticommutation relations in Eq.~\eqref{eqn:commu_psi_lattice}, all the operators in $S_x$ commute with each other and each has eigenvalues $0$ and $1$. The local fermion Hilbert space for a given $\alpha$ at one lattice site has dimension $2$ and the basis is specified by the eigenvalues of $\hat{\psi}^\dagger_\alpha(\hat{\boldsymbol x})\hat{\psi}_\alpha(\hat{\boldsymbol x})$. It can be mapped onto a qubit naturally, e.g., the eigenbasis with eigenvalue $n_\alpha$ maps onto the qubit state $|n_\alpha\rangle$ for $n_\alpha=0,1$. In order to maintain the anticommutation relations in Eq.~\eqref{eqn:commu_psi_lattice} when representing the Hamiltonian in this basis, one has to use either the Jordan-Wigner transformation or the Bravyi-Kitaev encoding to represent the fermion fields in terms of Pauli matrices. The former has an $O(\hat{V})$ overhead while the latter is more efficient with only an $O(\log \hat{L})$ overhead to implement an operator that is linear in $\psi$. We will provide more details in Sec.~\ref{sec:implementation}. The total number of qubits needed to encode the fermionic degrees of freedom is proportional to the lattice size, i.e., $4\hat{V}$.

\section{Bound on Qubit Cost to Represent Physical States}
\label{sec:qubit}
In this section, we will prove a bound for the total number of qubits needed for describing all physical states up to an energy $\hat{E}$ with an accuracy $\epsilon$, on a lattice of given size and fixed Hamiltonian parameters in lattice units. 
This type of bound analysis was first performed for quantum simulations of lattice scalar field theory~\cite{Jordan:2011ci} and later extended to the 2+1D SU(2) pure gauge theory in the irrep basis of the Kogut-Susskind Hamiltonian~\cite{Turro:2024pxu}, the latter of which can be easily generalized for SU($N_c$) non-Abelian gauge theories in higher dimensions in the irrep basis.
Equations~\eqref{eqn:Amax},~\eqref{eqn:Pimax}, and~\eqref{eqn:nA_final} are the new results obtained here for lattice QED in the Coulomb gauge.

We will focus on the gauge part since the total number of qubits needed for the fermion part is fixed to be $4\hat{V}$. We consider an arbitrary state with energy $\hat{E}$. Without truncation and digitization in the gauge field basis, it can be represented as
\bea
|\Psi \rangle &= \Bigg[\prod_{\hat{\boldsymbol x},i,\alpha} \int_{-\infty}^{\infty} {\rm d}\tilde{A}_i(\hat{\boldsymbol x}) \sum_{n_\alpha(\hat{\boldsymbol x})=0}^1 \Bigg] \bigotimes_{\hat{\boldsymbol x},i,\alpha} \big[ |\tilde{A}_i(\hat{\boldsymbol x}) \rangle  \otimes  |n_\alpha(\hat{\boldsymbol x})\rangle \big] \nn\\
&\quad \times \Psi\big[ \tilde{A}_1(\hat{\boldsymbol x}_1), \tilde{A}_2(\hat{\boldsymbol x}_1), \cdots, \tilde{A}_3(\hat{\boldsymbol x}_{\hat{V}}); n_1(\hat{\boldsymbol x}_1), n_2(\hat{\boldsymbol x}_1), \cdots, n_4(\hat{\boldsymbol x}_{\hat{V}}) \big] \,,
\eea
where the first term with the square bracket in the first line integrates and sums over all the field values, which is followed by the field basis states in the same line, and the last term denotes the state's wave function. If the gauge field values are truncated, we have an approximate representation of the state
\bea
|\Psi_{\rm cut} \rangle &= \Bigg[\prod_{\hat{\boldsymbol x},i,\alpha} \int_{-\tilde{A}_{\rm max}}^{\tilde{A}_{\rm max}} {\rm d}\tilde{A}_i(\hat{\boldsymbol x}) \sum_{n_\alpha(\hat{\boldsymbol x})=0}^1 \Bigg] \bigotimes_{\hat{\boldsymbol x},i,\alpha} \big[ |\tilde{A}_i(\hat{\boldsymbol x}) \rangle  \otimes  |n_\alpha(\hat{\boldsymbol x})\rangle \big] \nn\\
&\quad \times \Psi\big[ \tilde{A}_1(\hat{\boldsymbol x}_1), \tilde{A}_2(\hat{\boldsymbol x}_1), \cdots, \tilde{A}_3(\hat{\boldsymbol x}_{\hat{V}}); n_1(\hat{\boldsymbol x}_1), n_2(\hat{\boldsymbol x}_1), \cdots, n_4(\hat{\boldsymbol x}_{\hat{V}}) \big] \,.
\eea
Its overlap with the true state is
\bea
\langle \Psi | \Psi_{\rm cut} \rangle = \Bigg[\prod_{\hat{\boldsymbol x},i,\alpha} \int_{-\tilde{A}_{\rm max}}^{\tilde{A}_{\rm max}} {\rm d}\tilde{A}_i(\hat{\boldsymbol x}) \sum_{n_\alpha(\hat{\boldsymbol x})=0}^1 \Bigg] \Big| \Psi\big[ \{\tilde{A}_i(\hat{\boldsymbol x})\}; \{n_\alpha(\hat{\boldsymbol x})\} \big] \Big|^2 \,,
\eea
where we introduced an abbreviation for the arguments of the wave function. We let $P_{\rm out}(\hat{\bs x}, i) \equiv P(|\tilde{A}_i(\hat{\bs x})| > \tilde{A}_{\rm max})$ denote the probability of the state having a field value $\tilde{A}_i$ whose magnitude is greater than $\tilde{A}_{\rm max}$ at $\hat{\bs x}$.
Then through the probability of a union of events~\cite{Jordan:2011ci}, we find
\bea
\label{eqn:psi_psicut_bound}
\langle \Psi | \Psi_{\rm cut} \rangle \geq 1- \sum_{\hat{\bs x}, i} P_{\rm out}(\hat{\bs x}, i) \geq 1- 3\hat{V} \max_{\hat{\bs x}, i}P_{\rm out}(\hat{\bs x}, i) \,.
\eea
If we want an accuracy of $\epsilon$, i.e., $\langle \Psi | \Psi_{\rm cut} \rangle > 1-\epsilon$, we require
\bea
\label{eqn:Pmax_epsilon}
\max_{\hat{\bs x}, i}P_{\rm out}(\hat{\bs x}, i) \leq \frac{\epsilon}{3\hat{V}} \,.
\eea

The essential ingredient of the remaining proof is to use the Chebyshev's inequality to express $\max_{\hat{\bs x}, i}P_{\rm out}(\hat{\bs x}, i)$ in terms of the expected value of some field operator, which is bounded by the state energy $\hat{E}$, as done in the case of lattice scalar field theory~\cite{Jordan:2011ci}. 
The proof for lattice QED is more involved, since the bosonic fields are coupled with fermionic ones.
Another crucial difference here is that the analysis will only lead to a bound on $\tilde{A}_{\rm max}$ for physical states, i.e., the transverse components of the gauge fields. As explained in Sec.~\ref{H_equivalence} 
for the continuum theory and in Sec.~\ref{sec:lattice gauge invariance} for the lattice theory,
only the transverse gauge fields propagate and contribute to the electric and magnetic energies. The unphysical longitudinal gauge fields cannot be bounded by the physical quantity, the energy of the state, but this is fine since they are dynamically decoupled. We will explain the bound on $\tilde{A}_{\rm max}$ in Sec.~\ref{eqn:bound_Amax}. Putting a bound on $\delta\!\tilde{A}$ for physical states requires a similar analysis with the state expressed in the conjugate variable space, rather than the field space, which will be explained in Sec.~\ref{eqn:bound_deltaA}. In order to perform these analyses, we need to shift the Hamiltonian properly and decompose it into positive semidefinite parts, which we will do now.

\subsection{Positive Semidefinite Hamiltonian}
As explained above, we will focus on the transverse gauge fields, which are physical and propagating. This allows to drop the second term in $\hat{H}_\Pi$ in Eq.~\eqref{eqn:Hpi}, as well as the second term in $\hat{H}_A$ in Eq.~\eqref{eqn:HA}, both of which vanish for transverse gauge fields, as explained in Sec.~\ref{sec:lattice gauge invariance}. The remaining term, i.e., the first term in $\hat{H}_\Pi$ is already positive semidefinite. In order to reorganize the rest, e.g., the first term in $\hat{H}_A$ and $\hat{H}_I$, we introduce a discrete Fourier-like transform for $\hat{A}_i(\hat{\boldsymbol x})$
\begin{subequations}
\bea
\hat{A}_i(\hat{\boldsymbol x}) &= \sum_{\hat{\boldsymbol p}} \phi_{\hat{\boldsymbol{p}}}(\hat{\boldsymbol{x}}) \hat{A}_i(\hat{\boldsymbol p}) \,,\\
\hat{A}_i(\hat{\boldsymbol p}) &= \sum_{\hat{\boldsymbol x}} \phi_{\hat{\boldsymbol{p}}}(\hat{\boldsymbol{x}}) \hat{A}_i(\hat{\boldsymbol x}) \,,
\eea
\end{subequations}
and similarly for $\hat{J}^i(\hat{\boldsymbol x})$, where $\phi_{\hat{\boldsymbol{p}}}(\hat{\boldsymbol{x}}) = \phi_{\hat{p}_1}(\hat{x}_1) \phi_{\hat{p}_2}(\hat{x}_2) \phi_{\hat{p}_3}(\hat{x}_3)$ is given in Eq.~\eqref{eqn:phi_k(x)}.
Then we can write $\hat{H}_A+\hat{H}_I$ as
\bea
\label{eqn:HA_transform}
&\hat{H}_A +\hat{H}_I = 2 \sum_{\hat{\boldsymbol p},i,j} \sin^2\Big(\frac{\hat{p}_i}{2}\Big) \hat{A}_j(\hat{\boldsymbol p}) \hat{A}_j(\hat{\boldsymbol p}) - \sum_{\hat{\boldsymbol p},i} \hat{J}^i(\hat{\boldsymbol p}) \hat{A}_i(\hat{\boldsymbol p})  \nn\\
=\ & 2 \sum_{\hat{\boldsymbol p},j} \Big[\sum_i \sin^2\!\Big(\frac{\hat{p}_i}{2}\Big)\Big] \Big[ \hat{A}_j(\hat{\boldsymbol p}) - \frac{1}{4\sum_i \sin^2(\frac{\hat{p}_i}{2})} \hat{J}^j(\hat{\boldsymbol p}) \Big]^2 - \frac{1}{8} \sum_{\hat{\boldsymbol p},j} \frac{1}{\sum_i \sin^2(\frac{\hat{p}_i}{2})} \hat{J}^j(\hat{\boldsymbol p}) \hat{J}^j(\hat{\boldsymbol p}) \,.
\eea
Because both $\hat{A}_j(\hat{\bs x})$ and $\hat{J}_j(\hat{\bs x})$ are Hermitian operators and $\phi_{\hat{\boldsymbol{p}}}(\hat{\boldsymbol{x}})$ functions are real, both $\hat{A}_j(\hat{\bs p})$ and $\hat{J}_j(\hat{\bs p})$ are also Hermitian. We immediately see that each summand in the first term on the right-hand side of Eq.~\eqref{eqn:HA_transform}, i.e.,
\bea
\Big[\sum_i \sin^2\!\Big(\frac{\hat{p}_i}{2}\Big)\Big] \Big[ \hat{A}_j(\hat{\boldsymbol p}) - \frac{1}{4\sum_i \sin^2(\frac{\hat{p}_i}{2})} \hat{J}^j(\hat{\boldsymbol p}) \Big]^2 \,,
\eea
is positive semidefinite. The last term of Eq.~\eqref{eqn:HA_transform} can be combined with the Coulomb interaction term $\hat{H}_C$ in Eq.~\eqref{eqn:HC} to give
\bea
\label{eqn:HA_HC>=0}
&- \frac{1}{8} \sum_{\hat{\boldsymbol p},j} \frac{1}{\sum_i \sin^2(\frac{\hat{p}_i}{2})} \hat{J}^j(\hat{\boldsymbol p}) \hat{J}^j(\hat{\boldsymbol p}) - \frac{1}{2}\sum_{\hat{\boldsymbol x}, \hat{\boldsymbol y}}\hat{J}^0(\hat{\boldsymbol x})  G(\hat{\boldsymbol x}, \hat{\boldsymbol y}) \hat{J}^0(\hat{\boldsymbol y}) \nn\\
=&\ \frac{1}{8} \sum_{\hat{\boldsymbol p}} \frac{1}{\sum_i \sin^2(\frac{\hat{p}_i}{2})} \Big[ \hat{J}^0(\hat{\boldsymbol p})\hat{J}^0(\hat{\boldsymbol p}) - \sum_j \hat{J}^j(\hat{\boldsymbol p}) \hat{J}^j(\hat{\boldsymbol p})  \Big] \,,
\eea
where we have used the definition of $G(\hat{\boldsymbol x}, \hat{\boldsymbol y})$ in Eq.~\eqref{eqn:Green for discrete Laplacian}.
The physical meaning of $\hat{J}^\mu$ is the electromagnetic current density induced by the fermion $g\bar{\hat{\psi}}\gamma^\mu \hat{\psi}$ and is timelike for massive fermions and lightlike for massless ones. So $[\hat{J}^0(\hat{\boldsymbol p})]^2 - \sum_j [\hat{J}^j(\hat{\boldsymbol p})]^2 \geq 0$ and Eq.~\eqref{eqn:HA_HC>=0} is positive semidefinite.

The fermion Hamiltonian has negative eigenvalues, which is well known as the Dirac sea of negative-energy particles. We can shift the energy by a constant proportional to the lattice size such that all fermion states have nonnegative energies. To find this constant, we decompose the fermion fields in terms of the creation and annihilation operators in momentum space
\begin{subequations}
\label{eqn:quanti_psi}
\bea
\hat{\psi}(\hat{\boldsymbol x}) &= \frac{1}{\sqrt{\hat{V}}} \sum_{\hat{\boldsymbol p},s}\frac{1}{\sqrt{2\hat{E}_{\hat{\boldsymbol p}}}}\Big[ b_s(\hat{\boldsymbol p}) u_s(\hat{\boldsymbol p}) e^{i\hat{\boldsymbol p}\cdot \hat{\boldsymbol x}} + d_s^\dagger(\hat{\boldsymbol p}) v_s(\hat{\boldsymbol p}) e^{-i\hat{\boldsymbol p}\cdot \hat{\boldsymbol x}}\Big] \,,\\
\bar{\hat{\psi}}(\hat{\boldsymbol x}) &= \frac{1}{\sqrt{\hat{V}}}  \sum_{\hat{\boldsymbol p},s}\frac{1}{\sqrt{2\hat{E}_{\hat{\boldsymbol p}}}}\Big[ b_s^\dagger(\hat{\boldsymbol p}) \bar{u}_s(\hat{\boldsymbol p}) e^{-i\hat{\boldsymbol p}\cdot \hat{\boldsymbol x}} + d_s(\hat{\boldsymbol p}) \bar{v}_s(\hat{\boldsymbol p}) e^{i\hat{\boldsymbol p}\cdot \hat{\boldsymbol x}}\Big] \,,
\eea
\end{subequations}
where $s$ stands for spins and is summed over $\uparrow$ and $\downarrow$. $b$ and $b^\dagger$ are the annihilation and creation operators for fermions while $d$ and $d^\dagger$ are those for antifermions. They obey the standard anticommutation relations
\begin{subequations}
\bea
\{ b_s(\hat{\boldsymbol p}), b_{s'}^\dagger(\hat{\boldsymbol p}') \} &= \delta_{ss'}\delta_{\hat{\boldsymbol p}\hat{\boldsymbol p}'} \,,\\
\{ d_s(\hat{\boldsymbol p}), d_{s'}^\dagger(\hat{\boldsymbol p}') \} &= \delta_{ss'}\delta_{\hat{\boldsymbol p}\hat{\boldsymbol p}'} \,,
\eea
\end{subequations}
with all the other anticommutators vanishing. $u_s$ and $v_s$ are solutions to the Dirac equations 
\begin{subequations}
\bea
\Big( \gamma^0 \hat{E}_{\hat{\boldsymbol p}} - \gamma^i \sin\hat{p}^i - \hat{m} -2\hat{r} \sum_{j} \sin^2\frac{\hat{p}^j}{2} \Big) u_s(\hat{\boldsymbol p}) &= 0 \,,\\
\Big( \gamma^0 \hat{E}_{\hat{\boldsymbol p}} - \gamma^i \sin\hat{p}^i + \hat{m} +2\hat{r} \sum_{j} \sin^2\frac{\hat{p}^j}{2} \Big) v_s(\hat{\boldsymbol p}) &= 0 \,,
\eea
\end{subequations}
respectively, with the energy
\bea
\hat{E}_{\hat{\boldsymbol p}} = \sqrt{\sum_i\sin^2{\hat{p}_i} + \Big[\hat{m} + 2\hat{r} \sum_{j} \sin^2\!\Big(\frac{\hat{p}_j}{2}\Big)\Big]^2} \,.
\eea
Applying the field decomposition to the fermion Hamiltonian leads to
\bea
\hat{H}_f = \sum_{\hat{\boldsymbol p},s} \hat{E}_{\hat{\boldsymbol p} } \Big[ b_s^\dagger(\hat{\boldsymbol p}) b_s(\hat{\boldsymbol p}) + d_s^\dagger(\hat{\boldsymbol p}) d_s(\hat{\boldsymbol p}) - 1\Big] \,.
\eea
We can bound the constant term by using
\bea
|\hat{E}_{\hat{\boldsymbol p}}^2| \leq 3+\hat{m}^2 + 12\hat{m}\hat{r} + 36\hat{r}^2 \,.
\eea
It follows that if we shift the fermion Hamiltonian (and thus the total Hamiltonian) by the constant
\bea
\label{eqn:shift_const}
2\hat{V} \sqrt{3+\hat{m}^2 + 12\hat{m}\hat{r} + 36\hat{r}^2} \,,
\eea
the fermion Hamiltonian will become positive semidefinite. This argument is based on the occupation number basis in momentum space, in which the state with zero occupation number is defined to be annihilated by $b_s(\hat{\boldsymbol p})$ and $d_s(\hat{\boldsymbol p})$. It is worth emphasizing that we use the occupation number basis to calculate the shift constant that makes the fermion Hamiltonian positive semidefinite, but in practical simulations, the field basis specified by Eq.~\eqref{eqn:S_x} will be used.

\subsection{Bound on $\tilde{A}_{\rm max}$}
\label{eqn:bound_Amax}
After all the above preparation, we now prove a bound on $\tilde{A}_{\rm max}$ for transverse gauge fields. Because each term in the Hamiltonian is positive semidefinite after the constant shift given in Eq.~\eqref{eqn:shift_const}, we have
\bea
\label{eqn:bound_X_E'}
&\hat{E}' \equiv \hat{E} + 2\hat{V} \sqrt{3+\hat{m}^2 + 12\hat{m}\hat{r} + 36\hat{r}^2}  \geq \nn\\
&2 \langle \Psi | \sum_{\hat{\boldsymbol p},j} \Big[\sum_i \sin^2\!\Big(\frac{\hat{p}_i}{2}\Big)\Big] \Big[ \hat{A}_j(\hat{\boldsymbol p}) - \frac{1}{4\sum_i \sin^2(\frac{\hat{p}_i}{2})} \hat{J}^j(\hat{\boldsymbol p})\Big]^2   | \Psi \rangle \,.
\eea
With the Dirichlet boundary conditions, the momentum is bounded by the finite lattice size from below, $\hat{p}_i \geq \frac{\pi}{\hat{L}+1}$ as in Eq.~\eqref{eqn:discrete_momentum_Dirichlet}. The minimum value of $\sum_i \sin^2(\frac{\hat{p}_i}{2})$ is $3\sin^2\!\frac{\pi}{\hat{L}+1}$. Using this bound gives
\bea
\label{eqn:E'>A2}
\frac{\hat{E}'}{6\sin^2(\frac{\pi}{\hat{L}+1})} & \geq \langle \Psi | \sum_{\hat{\boldsymbol p},j} \Big[ \hat{A}_j(\hat{\boldsymbol p}) - \frac{1}{4\sum_i \sin^2(\frac{\hat{p}_i}{2})} \hat{J}^j(\hat{\boldsymbol p})\Big]^2  | \Psi \rangle \nn\\
&= \langle \Psi | \sum_{\hat{\boldsymbol x},j} \big\{\hat{A}_j(\hat{\boldsymbol x}) + \big[\nabla_{\hat{\boldsymbol{x}}}^{(L)}\big]^{-2} \hat{J}^j(\hat{\boldsymbol x}) \big\}^2 | \Psi \rangle \nn\\
& \geq \langle \Psi | \big\{\hat{A}_j(\hat{\boldsymbol x}) + \big[\nabla_{\hat{\boldsymbol{x}}}^{(L)}\big]^{-2} \hat{J}^j(\hat{\boldsymbol x}) \big\}^2 | \Psi \rangle \,,
\eea
where $\big[\nabla_{\hat{\boldsymbol{x}}}^{(L)}\big]^{-2}$ is understood as the Green's function in Eq.~\eqref{eqn:Green for discrete Laplacian}.
For notational simplicity, we introduce a new variable $X_j(\hat{\bs x})\equiv \hat{A}_j(\hat{\boldsymbol x}) + \big[\nabla_{\hat{\boldsymbol{x}}}^{(L)}\big]^{-2} \hat{J}^j(\hat{\boldsymbol x})$. The Chebyshev's inequality states
\bea
\label{eqn:X_chebyshev}
P(|X_i(\hat{\bs x}) - \mu_{X_i(\hat{\bs x})}| > \kappa \sigma_{X_i(\hat{\bs x})}) < \frac{1}{\kappa^2} \,,
\eea
where $\kappa>0$, $\mu_{X_i(\hat{\bs x})}$ and $\sigma_{X_i(\hat{\bs x})}$ are the mean and variance of the variable $X_i(\hat{\bs x})$. Using Proposition 2 of Ref.~\cite{Jordan:2011ci} gives
\begin{subequations}
\label{eqn:mu_sigma_X}
\bea
\mu_{X_i(\hat{\bs x})} &= \langle \Psi | X_i(\hat{\bs x}) | \Psi \rangle \leq \sqrt{\langle \Psi | [X_i(\hat{\bs x})]^2 | \Psi \rangle} \,,\\
\sigma_{X_i(\hat{\bs x})} &= \sqrt{ \langle \Psi | [X_i(\hat{\bs x}) - \mu_{X_i(\hat{\bs x})}]^2 | \Psi \rangle } \leq \sqrt{\langle \Psi | [X_i(\hat{\bs x})]^2 | \Psi \rangle} \,.
\eea
\end{subequations}
If we choose
\bea
\label{eqn:k_Xmax}
\kappa=\sqrt{\frac{3\hat{V}}{\epsilon}}\,, \quad X_{\rm max} = (\kappa+1)\sqrt{\frac{\hat{E}'}{6\sin^2(\frac{\pi}{\hat{L}+1})}} \approx \sqrt{\frac{\hat{E}'\hat{V}^{5/3}}{2\pi^2 \epsilon}} \,,
\eea
where we have assumed $\hat{L}\gg 1$ to obtain the approximation. We immediately see that Eqs.~\eqref{eqn:E'>A2},~\eqref{eqn:X_chebyshev}, and~\eqref{eqn:mu_sigma_X} lead to
\bea
P(|X_i(\hat{\boldsymbol x})|>X_{\rm max}) < \frac{\epsilon}{3\hat{V}} \,.
\eea
This is a bound on $X_{\rm max}$. To obtain a bound on $\tilde{A}_{\rm max}$, we use
\bea
| \tilde{A}_i(\hat{\boldsymbol x}) | \leq | X_i(\hat{\boldsymbol x}) | + \big| \big[\nabla_{\hat{\boldsymbol{x}}}^{(L)}\big]^{-2} \hat{J}^i(\hat{\boldsymbol x}) \big| \,,
\eea
and choose 
\bea
\label{eqn:Amax_raw}
\tilde{A}_{\rm max} = X_{\rm max} + \max_{\hat{\boldsymbol x},i} \big| \big[\nabla_{\hat{\boldsymbol{x}}}^{(L)}\big]^{-2} \hat{J}^i(\hat{\boldsymbol x}) \big| \,.
\eea
Then we can show
\bea
P(|\tilde{A}_i(\hat{\boldsymbol x})| \leq \tilde{A}_{\rm max}) &\geq 
P\Big( |X_i(\hat{\boldsymbol x})| + \big| \big[\nabla_{\hat{\boldsymbol{x}}}^{(L)}\big]^{-2} \hat{J}^i(\hat{\boldsymbol x}) \big| \leq X_{\rm max} + \max_{\hat{\boldsymbol x},i} \big| \big[\nabla_{\hat{\boldsymbol{x}}}^{(L)}\big]^{-2} \hat{J}^i(\hat{\boldsymbol x}) \big| \Big) \nn\\
&\geq P( |X_i(\hat{\boldsymbol x})| \leq X_{\rm max} ) \geq 1-\frac{\epsilon}{3\hat{V}} \,.
\eea
In other words,
\bea
P(|\tilde{A}_i(\hat{\boldsymbol x})| > \tilde{A}_{\rm max}) < \frac{\epsilon}{3\hat{V}} \,,
\eea
and from Eq.~\eqref{eqn:psi_psicut_bound} we can conclude $\langle \Psi|\Psi_{\rm cut} \rangle > 1-\epsilon$.

Therefore, the last step to bound $\tilde{A}_{\rm max}$ is obtain $\max_{\hat{\boldsymbol x},i} \big| \big[\nabla_{\hat{\boldsymbol{x}}}^{(L)}\big]^{-2} \hat{J}^i(\hat{\boldsymbol x}) \big|$. In our lattice setup, no particular spatial direction is preferred. So it suffices to choose one spatial direction $i$ for the analysis of finding the maximum. We choose the $z$ direction, i.e., $i=3$. In the Weyl representation of gamma matrices, we have
\bea
\hat{J}^3 = g(-\hat{\psi}^\dagger_1 \hat{\psi}_1 + \hat{\psi}^\dagger_2 \hat{\psi}_2 + \hat{\psi}^\dagger_3 \hat{\psi}_3 - \hat{\psi}^\dagger_4 \hat{\psi}_4) \,,
\eea
where we have omitted the position dependence in the fields, since it is clear that we now focus on one spatial point. The local fermion Hilbert space at this position is of dimension 16 and spanned by states of the form
\bea
|n_1,n_2,n_3,n_3\rangle = (\hat{\psi}_1^\dagger)^{n_1} (\hat{\psi}_2^\dagger)^{n_2} (\hat{\psi}_3^\dagger)^{n_3} (\hat{\psi}_4^\dagger)^{n_4} |0,0,0,0\rangle \,,
\eea
where $n_i\in\{0,1\}$.
They are also eigenstates of $\hat{J}^3$ with the eigenvalues $g(-n_1+n_2+n_3-n_4)$, i.e.,
\bea
\hat{J}^3 |n_1,n_2,n_3,n_3\rangle = g(-n_1+n_2+n_3-n_4) |n_1,n_2,n_3,n_3\rangle \,.
\eea
The maximal eigenvalue magnitude is $2g$, so we conclude $\max_{\hat{\bs x},i}|\hat{J}^i(\hat{\bs x})|  = \max_{\hat{\bs x}}|\hat{J}^3(\hat{\bs x})| = 2g$. 
Together with the fact that the maximum value of $\big|\big[\nabla_{\hat{\boldsymbol{x}}}^{(L)}\big]^{-2}\big|$ is given by the inverse of the minimum eigenvalue of $\big|\big[\nabla_{\hat{\boldsymbol{x}}}^{(L)}\big]^{2}\big|$, which is $12\sin^2\!\frac{\pi}{\hat{L}+1}$, we find
\bea
\label{eqn:max_nablaJ}
\max_{\hat{\boldsymbol x},i} \big|\big[\nabla_{\hat{\boldsymbol{x}}}^{(L)}\big]^{-2} \hat{J}^i(\hat{\boldsymbol x}) \big| \leq \frac{2g}{12} \sin^{-2}\!\frac{\pi}{\hat{L}+1} \approx \frac{g\hat{L}^2}{6\pi^2} \,.
\eea

Combining Eqs.~\eqref{eqn:k_Xmax},~\eqref{eqn:Amax_raw} and~\eqref{eqn:max_nablaJ}, we obtain a bound on $\tilde{A}_{\rm max}$ for the physical transverse gauge fields
\bea
\label{eqn:Amax}
\tilde{A}_{\rm max} = \sqrt{\frac{\hat{E}'\hat{V}^{5/3}}{2\pi^2 \epsilon}} + \frac{g\hat{V}^{2/3}}{6\pi^2} \,.
\eea

\subsection{Bound on $\delta\!\tilde{A}$}
\label{eqn:bound_deltaA}
From Eq.~\eqref{eqn:commu_APi_lattice}, we can write the conjugate variable as a functional derivative in the field space
\bea
\label{eqn:Pi_as_ddA}
\hat{\Pi}_i(\bs x) = -i\frac{\delta}{\delta \hat{A}_i(\bs x)} \,.
\eea
Instead of the field space, a state can also be represented in the conjugate variable space defined by 
\bea
\hat{\Pi}_i(\bs x)|\tilde{\Pi}_j(\bs y) \rangle = \delta_{ij}\delta_{{\bs x}{\bs y}}\tilde{\Pi}_j(\bs y)|\tilde{\Pi}_j(\bs y) \rangle \,.
\eea
According to Proposition 1 in Ref.~\cite{Jordan:2011ci}, the field space and the conjugate variable space at a given position can be swapped by a local Fourier transform, which further leads to
\bea
\label{eqn:Pimax_deltaA}
\tilde{\Pi}_{\rm max} = \frac{\pi}{\delta\!\tilde{A}} \,.
\eea
This gives us a way to bound $\delta\!\tilde{A}$ by using the energy of the state.

Only the first term in Eq.~\eqref{eqn:Hpi} contributes to the electric energy, since we focus on the physical transverse fields. An inequality similar to Eq.~\eqref{eqn:E'>A2} gives
\bea
\hat{E}' \geq \langle \Psi | \frac{1}{2}\hat{\Pi}^2_i(\hat{\boldsymbol x})  | \Psi \rangle \,,
\eea
for any $i$ and $\hat{\bs x}$. Using analogs of Eqs.~\eqref{eqn:X_chebyshev} and~\eqref{eqn:mu_sigma_X}, we find that if we choose
\bea
\label{eqn:Pimax}
\tilde{\Pi}_{\rm max} = \sqrt{\frac{6\hat{E}'\hat{V}}{\epsilon}} \,,
\eea
we will have
\bea
P(|\tilde{\Pi}_i(\hat{\boldsymbol x})|>\tilde{\Pi}_{\rm max}) < \frac{\epsilon}{3\hat{V}} \,,
\eea
and the approximate state's fidelity bounded as
$\langle \Psi | \Psi_{\rm cut} \rangle \geq 1-\epsilon$.

\subsection{Total Cost of Qubits}
We conclude this section by writing down the expression for the total cost of qubits to represent the fermion and physical transverse gauge fields of a state with an infidelity $\epsilon$. By using Eqs.~\eqref{eqn:nA},~\eqref{eqn:Amax},~\eqref{eqn:Pimax_deltaA}, and~\eqref{eqn:Pimax}, we conclude that the number of qubits needed to represent the transverse gauge fields at one lattice site is
\bea
\label{eqn:nA_final}
n_A \approx \log_2\left( \frac{2\sqrt{3}\hat{E}'\hat{V}^{4/3}}{\pi^2\epsilon} + \frac{\sqrt{2}g \hat{E}'^{1/2} \hat{V}^{7/6}}{\sqrt{3}\pi^3\epsilon^{1/2}} \right) \,.
\eea
The qubit cost for the fermion field per lattice site is $4$. Finally, the total cost is given by
\bea
\label{eqn:nA_V}
3n_A\hat{V}+4\hat{V} \approx 3\hat{V}\log_2\left( \frac{2\sqrt{3}\hat{E}'\hat{V}^{4/3}}{\pi^2\epsilon} + \frac{\sqrt{2}g \hat{E}'^{1/2} \hat{V}^{7/6}}{\sqrt{3}\pi^3\epsilon^{1/2}} \right) + 4\hat{V} \,.
\eea

\section{Quantum Algorithm for Real-Time Simulation}
\label{sec:real-time}
\subsection{Trotterization Error}
We will use Trotterization to implement the Hamiltonian evolution. To this end, we need to decompose the Hamiltonian into different pieces such that within the same piece, every term commutes with each other. Because of Eq.~\eqref{eqn:commu_A_lattice}, we immediately see that $\hat{H}_\Pi$ in Eq.~\eqref{eqn:Hpi} and $\hat{H}_A$ in Eq.~\eqref{eqn:HA} are two such pieces. For the terms involving fermion fields, we first note that in the Weyl representation of the gamma matrices, $\hat{H}_I+\hat{H}_C+\hat{H}_f$ has no terms of the form $\psi^\dagger_1(\bs x)\psi_4(\bs y)$ or $\psi^\dagger_2(\bs x)\psi_3(\bs y)$. Furthermore, we have
\begin{subequations}
\bea
[\hat{\psi}^\dagger_\alpha(\boldsymbol x)\hat{\psi}_\alpha(\boldsymbol x), \hat{\psi}^\dagger_\beta(\boldsymbol y)\hat{\psi}_\beta(\boldsymbol y) ] & = 0\,,\quad \forall \alpha,\beta,{\boldsymbol x}, {\boldsymbol y} \,,\\
[\hat{\psi}^\dagger_\alpha(\boldsymbol x)\hat{\psi}_\beta(\boldsymbol x), \hat{\psi}^\dagger_\gamma(\boldsymbol y)\hat{\psi}_\rho(\boldsymbol y) ] & = 0 \,, \quad \forall {\boldsymbol x}\neq {\boldsymbol y} \,,
\eea
\end{subequations}
where in the first line there is no summation over $\alpha$ or $\beta$. Using these vanishing commutators, we find that the Hamiltonians that involve fermions can be decomposed into 15 pieces, each of which only consists of commuting terms. These terms are of the forms
\begin{subequations}
\bea
\hat{H}_{11i}^{e}&: \{ \hat{\psi}_1^\dagger(\hat{\boldsymbol x})\hat{\psi}_1(\hat{\boldsymbol x}+\hat{i}),\,\hat{\psi}_2^\dagger(\hat{\boldsymbol x})\hat{\psi}_2(\hat{\boldsymbol x}+\hat{i}),\,\hat{\psi}_3^\dagger(\hat{\boldsymbol x})\hat{\psi}_3(\hat{\boldsymbol x}+\hat{i}),\,\hat{\psi}_4^\dagger(\hat{\boldsymbol x})\hat{\psi}_4(\hat{\boldsymbol x}+\hat{i})\ |\ \forall \hat{\boldsymbol x}, \hat{x}_i\ {\rm even} \} \,,\\
\hat{H}_{12i}&: \{ \hat{\psi}_1^\dagger(\hat{\boldsymbol x})\hat{\psi}_2(\hat{\boldsymbol x}+\hat{i}),\, \hat{\psi}_1^\dagger(\hat{\boldsymbol x})\hat{\psi}_3(\hat{\boldsymbol x}+\hat{i}),\,
\hat{\psi}_4^\dagger(\hat{\boldsymbol x})\hat{\psi}_2(\hat{\boldsymbol x}+\hat{i}),\,
\hat{\psi}_4^\dagger(\hat{\boldsymbol x})\hat{\psi}_3(\hat{\boldsymbol x}+\hat{i})\ |\ \forall \hat{\boldsymbol x} \} \,,\\
\hat{H}_{21i}&: \{ \hat{\psi}_2^\dagger(\hat{\boldsymbol x})\hat{\psi}_1(\hat{\boldsymbol x}+\hat{i}),\,
\hat{\psi}_3^\dagger(\hat{\boldsymbol x})\hat{\psi}_1(\hat{\boldsymbol x}+\hat{i}),\,
\hat{\psi}_2^\dagger(\hat{\boldsymbol x})\hat{\psi}_4(\hat{\boldsymbol x}+\hat{i}),\,
\hat{\psi}_3^\dagger(\hat{\boldsymbol x})\hat{\psi}_4(\hat{\boldsymbol x}+\hat{i})\ |\ \forall \hat{\boldsymbol x} \} \,,\\
\hat{H}_{11}^{s}&: \{ \hat{\psi}_1^\dagger(\hat{\boldsymbol x})\hat{\psi}_1(\hat{\boldsymbol x}),\,\hat{\psi}_2^\dagger(\hat{\boldsymbol x})\hat{\psi}_2(\hat{\boldsymbol x}),\,\hat{\psi}_3^\dagger(\hat{\boldsymbol x})\hat{\psi}_3(\hat{\boldsymbol x}),\,\hat{\psi}_4^\dagger(\hat{\boldsymbol x})\hat{\psi}_4(\hat{\boldsymbol x})\ |\ \forall \hat{\boldsymbol x} \} \,,\\
\hat{H}_{12}^{s}&: \{ \hat{\psi}_1^\dagger(\hat{\boldsymbol x})\hat{\psi}_2(\hat{\boldsymbol x}),\,\hat{\psi}_1^\dagger(\hat{\boldsymbol x})\hat{\psi}_3(\hat{\boldsymbol x}),\,\hat{\psi}_4^\dagger(\hat{\boldsymbol x})\hat{\psi}_2(\hat{\boldsymbol x}),\,\hat{\psi}_4^\dagger(\hat{\boldsymbol x})\hat{\psi}_3(\hat{\boldsymbol x})\ |\ \forall \hat{\boldsymbol x} \} \,,\\
\hat{H}_{21}^{s}&: \{ \hat{\psi}_2^\dagger(\hat{\boldsymbol x})\hat{\psi}_1(\hat{\boldsymbol x}),\,\hat{\psi}_3^\dagger(\hat{\boldsymbol x})\hat{\psi}_1(\hat{\boldsymbol x}),\,\hat{\psi}_2^\dagger(\hat{\boldsymbol x})\hat{\psi}_4(\hat{\boldsymbol x}),\,\hat{\psi}_3^\dagger(\hat{\boldsymbol x})\hat{\psi}_4(\hat{\boldsymbol x})\ |\ \forall \hat{\boldsymbol x} \} \,.
\eea
\end{subequations}
In the first line, the superscript $e$ indicates $\hat{x}_i$ taking even sites. They correspond to three pieces of the Hamiltonians. By replacing $e$ with $o$, which means $\hat{x}_i$ taking odd sites, we have another three pieces. The second and third lines form another six pieces. The last three pieces are made of the forms in the last three lines.

Taking into account $\hat{H}_\Pi$ and $\hat{H}_A$, we have 17 pieces in total, labeled by $\hat{H}_i$, $i\in\{1,2,\cdots,17\}$. First-order Trotterization gives~\cite{trotter1959product,Suzuki:1976be}
\bea
e^{-i\hat{H}\hat{t}} = \prod_{k=0}^{N_t-1} \prod_{i=1}^{17} e^{-i\hat{H}_i\Delta \hat{t}} +O\bigg(\sum_{i,j}\big|\big| [\hat{H}_i,\hat{H}_j] \big|\big|  \frac{\hat{t}^2}{N_t}\bigg) \,,
\eea
where the Trotter step size is $\Delta \hat{t}=\hat{t}/N_t$.
The Trotterization error is proportional to $\hat{t}^2/N_t$, with the proportional constant given by the sum of many operator norms $||[\hat{H}_i,\hat{H}_j]||$. We now estimate their scalings. Instead of using concrete expressions of $\hat{H}_i$, $i\in\{1,2,\cdots,17\}$, we can use Eq.~\eqref{eqn:latticeH} for the estimate. We note that the fermion operators $\hat{\psi}_\alpha(\bs x)$ and $\hat{\psi}_\alpha^\dagger(\bs x)$ just flip the fermion state between 0 and 1, so their norm is order one. The norm of the commutator between the fermion Hamiltonians $\hat{H}_C$ and $\hat{H}_f$ scale as
\bea
\label{eqn:normHcHf}
\big|\big| [\hat{H}_C, \hat{H}_f] \big|\big| = O(g^2\hat{V}^{8/3}) + O(g^2\hat{m}\hat{V}^{8/3}) + O(g^2\hat{r}\hat{V}^{8/3}) \,.
\eea
The norms of the commutators involving $\hat{H}_I$ take the forms
\begin{subequations}
\label{eqn:normHI}
\bea
\big|\big| [\hat{H}_I, \hat{H}_C] \big|\big| &= O(g^3\hat{V}^{8/3} \tilde{A}_{\rm max}) \,,\\
\big|\big| [\hat{H}_I, \hat{H}_f] \big|\big| &= O(g\hat{V}\tilde{A}_{\rm max}) + O(g\hat{m}\hat{V}\tilde{A}_{\rm max}) + O(g\hat{r}\hat{V}\tilde{A}_{\rm max}) \,,\\
\big|\big| [\hat{H}_I, \hat{H}_\Pi] \big|\big| &\sim \sum_{\hat{\boldsymbol x},i,j} \Big|\Big| g\bar{\hat{\psi}}(\hat{\boldsymbol x}) \gamma^i \hat{\psi}(\hat{\boldsymbol x}) \Big( \delta_{ij} - \partial_{\hat{x}_i}^{(L+)} \big[\nabla_{\hat{\boldsymbol{x}}}^{(L)}\big]^{-2} \partial_{\hat{x}_j}^{(L-)} \Big) \hat{\Pi}_j(\hat{\boldsymbol x}) \Big|\Big|  = O(g\hat{V}\tilde{\Pi}_{\rm max}) \,,
\eea
\end{subequations}
where in the last line we used a simple notation $\big[\nabla_{\hat{\boldsymbol{x}}}^{(L)}\big]^{-2}$ for the Green's function, see Eq.~\eqref{eqn:Hpi} for the full expression in position space. The final nonvanishing commutator is $[\hat{H}_\Pi, \hat{H}_A]$, the norm of which can be estimated as
\bea
\label{eqn:normHA}
\big|\big| [\hat{H}_\Pi, \hat{H}_A] \big|\big| \sim \sum_{\hat{\boldsymbol x},i,j} \big|\big|  \hat{\Pi}_i(\hat{\boldsymbol x}) \big( \big[\nabla_{\hat{\boldsymbol{x}}}^{(L)}\big]^2 \delta_{ij} - \partial_{\hat{x}_i}^{(L+)} \partial_{\hat{x}_j}^{(L-)} \big) \hat{A}_j(\hat{\boldsymbol x}) \big|\big| = O(\hat{V} \tilde{\Pi}_{\rm max} \tilde{A}_{\rm max} ) \,.
\eea
If we want the Trotterized time evolution till time $t$ to have an error $\epsilon$ at most, we require the number of Trotter steps to be at least
\bea
\label{eqn:Nt_scaling}
N_t \sim \sum_{i,j}\big|\big| [\hat{H}_i,\hat{H}_j] \big|\big| \frac{\hat{t}^2}{\epsilon} \,.
\eea
From Eqs.~\eqref{eqn:normHcHf},~\eqref{eqn:normHI},~\eqref{eqn:normHA},~\eqref{eqn:Amax}, and~\eqref{eqn:Pimax}, we conclude $N_t$ scales polynomially with the accuracy, time length, state energy, lattice size, and Hamiltonian parameters in lattice units.

\subsection{Implementation Cost for Each Trotter Step}
\label{sec:implementation}
We now discuss how to implement each piece $\hat{H}_i$ in the Trotterized Hamiltonian evolution. First we discuss the evolution driven by $\hat{H}_A$. The computational basis for gauge degrees of freedom is the field basis in position space, in which $\hat{H}_A$ is diagonal. It only induces a phase rotation. Classically it takes $O(\hat{V})$ to compute this phase as Eq.~\eqref{eqn:HA} contains a summation over the volume. In quantum computing, it takes $O(\hat{V})$ gates to implement this phase rotation, via, e.g., the phase kickback method, as discussed in the case of scalar field theory~\cite{Jordan:2011ci}. One can also decompose the phase rotation in terms of tensor products of identity and Pauli Z operators, which can be implemented with standard methods in quantum computing. A concrete construction can be found in Sec.~\ref{sec:comparison} that costs $O(n_A^2\hat{V})$ gates, where $n_A$ is the number of qubits needed to describe the local gauge degrees of freedom at one site and its value is estimated in Eq.~\eqref{eqn:nA_final}. Phase rotations at different sites can be implemented in parallel.

Next we discuss the evolution induced by $\hat{H}_\Pi$, which is diagonal in the conjugate variable space. As can be seen from Eq.~\eqref{eqn:Pi_as_ddA} and discussions there, the conjugate variable space at one spatial point is related to the field space at the same location via a Fourier transform. Using the quantum Fourier transform algorithm, we can efficiently convert between the field space and the conjugate variable space, which is similar to the case of scalar field theory~\cite{Jordan:2011ci}. The cost of the quantum Fourier transform is $O(n_A^2)$. Once we convert to the conjugate variable space, the evolution $e^{-i\hat{H}_\Pi \Delta \hat{t}}$ is just a phase rotation. Classically it takes $O(\hat{V}^2)$ to compute this phase [note that the second term in $\hat{H}_\Pi$ in Eq.~\eqref{eqn:Hpi} contains two spatial summations while $\hat{H}_A$ only has one]. Transforming back to the field space after the $e^{-i\hat{H}_\Pi \Delta \hat{t}}$ phase rotation takes another $O(n_A^2)$ gates for the quantum Fourier transform.

Finally, we discuss the implementation of the Hamiltonian pieces involving fermions. As mentioned earlier, to maintain the anticommutation relation of fermion fields~\eqref{eqn:commu_psi_lattice}, one has to apply the Jordan-Wigner transformation or the Bravyi-Kitaev encoding. Here we use the Jordan-Wigner transformation. In 3D spatial lattice, one can find a path going through every point without repeating, via, e.g., a snake-shape path. This path defines a map
\bea
\hat{\boldsymbol x}, \alpha \mapsto l_\alpha(\hat{\boldsymbol x}) \,,
\eea
where $\alpha$ is the fermion index and $l_\alpha(\hat{\boldsymbol x})$ is an integer labeling the position along the path. As position changes and $\alpha$ changes from 1 to 4, $l_\alpha(\hat{\boldsymbol x})$ increases. The increment is always $1$. Specifically, at the $n$th point on the path $\hat{\boldsymbol x}_n$, $l_\alpha(\hat{\boldsymbol x}_n)=4n+\alpha$. With this map, we can implement an arbitrary two-fermion-field operator as
\bea
\label{eqn:JW}
\psi^\dagger_\alpha(\hat{\boldsymbol x}) \psi_\beta(\hat{\boldsymbol y}) \to \sigma^-_{l_\alpha(\hat{\boldsymbol x})} \otimes \sigma^z_{l_\alpha(\hat{\boldsymbol x})+1} \otimes \sigma^z_{l_\alpha(\hat{\boldsymbol x})+2} \otimes \cdots \otimes \sigma^z_{l_\beta(\hat{\boldsymbol y})-1} \otimes \sigma^+_{l_\beta(\hat{\boldsymbol y})} \,,
\eea
where $\sigma^{\pm} = (\sigma^x\pm i\sigma^y)/2$ and $\sigma^x$, $\sigma^y$, and $\sigma^z$ are Pauli matrices. The fermion state with occupation number zero (one) corresponds to the eigenstate of $\sigma^z$ with eigenvalue $1$ ($-1$). The subscripts of the Pauli operators indicate where they act on. The Pauli Z operators in the middle are overheads for maintaining the fermion anticommutation relations. In the Hamiltonians $\hat{H}_I$, $\hat{H}_C$, and $\hat{H}_f$, we only have cases where $\bs y$ is at most one site away from $\bs x$. Therefore, the worst case of the overhead in a 3D snake-shape path happens between two layers of snake planes, corresponding to $O(\hat{L}^2)$ Pauli Z operators in the middle. The two fermion fields in the summand in $\hat{H}_I$ are at the same site, so the overhead is $O(1)$. The gauge field $\hat{A}_i$ in $\hat{H}_I$ is diagonal in the gauge field basis and thus can be decomposed into tensor products of identity and Pauli Z operators, using $O({n_A})$ terms (see Sec.~\ref{sec:comparison}). $\hat{H}_I$ sums over $O(\hat{V})$ terms so the total cost for its implementation is $O(n_A\hat{V})$. According to Eq.~\eqref{eqn:nA_final}, ${n_A}$ scales logarithmically with energy, volume, accuracy and Hamiltonian parameters in lattice units. The Hamiltonian piece $\hat{H}_C$ has four fermion fields, forming two pairs. Each pair is at the same location with the same index and thus has no overhead. Thus, $\hat{H}_C$ can be implemented with $O(\hat{V}^2)$ gates. Finally, in $\hat{H}_f$, the worst overhead is $O(\hat{L}^2)$ as mentioned above. Each summation in $\hat{H}_f$ contains $O(\hat{V})$ terms. So it is expected that the cost for implementing $\hat{H}_f$ scales as $O(\hat{V}^{5/3})$. More efficient implementation of fermion operators can be found in Ref.~\cite{Halimeh:2025ivn}.

All in all, the cost for the implementation of each piece in the Trotterized Hamiltonian evolution scales polynomially with energy, volume, accuracy, and Hamiltonian parameters in lattice units. Together with Eq.~\eqref{eqn:Nt_scaling}, we see that the total cost for real-time simulation scales polynomially.

\section{Comparison with Previous Work}
\label{sec:comparison}
The previous work~\cite{Li:2024ide} considered simulating lattice QED in the Coulomb gauge by using the occupation basis in momentum space for the gauge fields. Here we show the field basis in position space used in this work is polynomially more efficient in terms of single-qubit rotation and two-qubit controlled-not (CNOT) gate counts for simulating time evolution. The number of CNOT gates is a crucial resource for quantum simulations in the noisy intermediate-scale quantum era while the number of T gates is more important in fault-tolerant quantum computing, the latter of which is related to the number of single-qubit rotation by a factor of $10$--$50$~\cite{Halimeh:2024bth}, see also Refs.~\cite{Kliuchnikov:2014pzk,Campbell:2020wqh}.
Since the fermion fields are labeled in position space in both studies, the gate count for the fermion part of the Hamiltonian is the same. Therefore, we will only compare the gate counts for the gauge field part, i.e., the free part $\hat{H}_{\Pi}+\hat{H}_A$ in Eqs.~\eqref{eqn:Hpi} and~\eqref{eqn:HA} and the interaction part $\hat{H}_I$ in Eq.~\eqref{eqn:HI}. 

To estimate the gate count when using the field basis in position space, we utilize a well-known result that the field operator $\hat{A}(\hat{\bs x})$ (we omit the spatial index $i$ for notational simplicity here) under $2^{n_A}$ digitizations can be decomposed in terms of Pauli operators as~\cite{Klco:2018zqz,Halimeh:2024bth}
\bea
\label{eqn:Adecomp}
\hat{A}(\hat{\bs x}) = -\frac{1}{2}\delta\! \tilde{A} \sum_{j=0}^{n_A-1} 2^j \sigma^z_j(\hat{\bs x}) \,,
\eea
where $\sigma^z_j(\hat{\bs x})$ denotes the Pauli Z operator acting on the $j$th qubit representing the gauge field $\hat{A}(\hat{\bs x})$ at position $\hat{\bs x}$. Then we see that the Hamiltonian $\hat{H}_A$ in Eq.~\eqref{eqn:HA} can be written as a sum of many terms of the form $\sigma^z_{j_1}(\hat{\bs x}_1) \otimes \sigma^z_{j_2}(\hat{\bs x}_2) $, where $\hat{\bs x}_2$ can be $\hat{\bs x}_1$ or its nearest neighbors. The number of such terms is $O(n_A^2\hat{V})$. We conclude that for one Trotterized time evolution driven by $\hat{H}_A$, the gate counts for single-qubit rotation and CNOT are both $O(n_A^2\hat{V})$, since one time evolution step driven by $\sigma^z_{j_1}(\hat{\bs x}_1) \otimes \sigma^z_{j_2}(\hat{\bs x}_2) $ can be realized by one single-qubit rotation and two CNOT gates.

The conjugate field operator $\hat{\Pi}(\hat{\bs x})$ has a similar decomposition in the conjugate variable basis, which is related to the field basis by a discrete Fourier transform locally at the same position. The quantum Fourier transform algorithm requires $O(n_A^2)$ single-qubit rotation and CNOT gates per lattice site (a more efficient algorithm has been constructed~\cite{Nam:2019sox}). In the conjugate variable basis, $\hat{\Pi}(\hat{\bs x})$ is diagonal and the Hamiltonian $\hat{H}_{\Pi}$ in Eq.~\eqref{eqn:Hpi} can be written as a sum of many $\sigma^z_{j_1}(\hat{\bs x}_1) \otimes \sigma^z_{j_2}(\hat{\bs x}_2)$ terms. Their number is $O(n_A^2\hat{V}^2)$, since the second term in Eq.~\eqref{eqn:Hpi} contains two summations over the volume. Therefore, simulating one Trotter step driven by $\hat{H}_{\Pi}$ requires $O(n_A^2\hat{V}) + O(n_A^2\hat{V}^2) = O(n_A^2\hat{V}^2)$ single-qubit rotation and CNOT gates, where we see the gate cost for implementing the quantum Fourier transform is negligible.

Finally, to estimate the gate count for the time evolution driven by $\hat{H}_I$ in Eq.~\eqref{eqn:HI}, we note that each summand is of the form $\psi^\dagger_\alpha(\hat{\bs x}) \psi_\beta(\hat{\bs x}) \hat{A}(\hat{\bs x})$ and can be decomposed as a sum of tensor products of Pauli matrices by using Eqs.~\eqref{eqn:JW} and~\eqref{eqn:Adecomp}. Each tensor product at most involves five Pauli matrices, since the two fermion fields are at the same position. One Trotterized time evolution driven by a tensor product of five Pauli matrices can be implemented by one single-qubit rotation and eight CNOT gates, plus some single-qubit Hadamard and phase gates to convert Pauli X and Y gates to Pauli Z gates. Thus, one Trotter step driven by $\hat{H}_I$ at most requires $O(n_A\hat{V})$ single-qubit rotation and CNOT gates.

In a nutshell, the single-qubit rotation and CNOT gate costs for implementing one Trotter step in the time evolution of the gauge field Hamiltonian are summarized in Table~\ref{tab:gate_cost}, compared with those estimated in the previous work~\cite{Li:2024ide} that uses the occupation basis in momentum space. 
$n_A$ is given in Eq.~\eqref{eqn:nA_final} and we see it scales logarithmically with the coupling $g$, the energy $\hat{E}$, the volume $\hat{V}$, and the accuracy $\epsilon$. 
Equation~(11) of Ref.~\cite{Li:2024ide} leads to an estimate of the photon occupation number cutoff $\Lambda=O(g\hat{V}^3/\epsilon)+O(g^2\hat{V}^4/\epsilon)+O(\hat{V}^{7/3}/\epsilon)+O(\hat{E}\hat{V}^{4/3}/\epsilon)$, by which we mean that in different asymptotic parameter regimes, at least one term dominates.
Although for the implementation of one Trotter step driven by the free gauge Hamiltonian $\hat{H}_{\Pi}+\hat{H}_A$, the field basis in position space used in this work is less efficient by a factor of $O(n_A^2\hat{V}^2)$, it is more efficient in the implementation of the interaction Hamiltonian evolution driven by $\hat{H}_I$, by a factor of $O(\hat{V}\Lambda\log_2\Lambda)$, if we take $n_A\sim \log_2\Lambda$ roughly. Considering the total gate costs for both Hamiltonian terms together, we conclude that using the field basis in position space is polynomially more efficient. The total cost for each gate type is reduced by a factor of $O(\Lambda)$. For a modest size lattice $\hat{L}=10$ and a modest accuracy $\epsilon=10\%$, $\Lambda$ is at least on the order of $10^8$, estimated from the $\hat{V}^{7/3}/\epsilon$ term that is independent of the coupling and energy.

\begin{table}[h]
\centering
\begin{tabular}{|c|c|c|c|c|}
    \hline
    \multicolumn{2}{|c|}{Gate counts} & $\hat{H}_{\Pi}+\hat{H}_A$ & $\hat{H}_{I}$ & $\hat{H}_{\Pi}+\hat{H}_A+\hat{H}_{I}$\\
    \hline
    \multirow{2}{*}{Single-qubit rotation} & This work & $O(n_A^2\hat{V}^2)$ & $O(n_A\hat{V})$ & $O(n_A^2\hat{V}^2)$ \\
    \cline{2-5} & Ref.~\cite{Li:2024ide} & $O(\hat{V}\log_2\Lambda)$ & $O[\hat{V}^2\Lambda(\log_2\Lambda)^2]$ & $O[\hat{V}^2\Lambda(\log_2\Lambda)^2]$ \\
    \hline
    \multirow{2}{*}{CNOT} & This work & $O(n_A^2\hat{V}^2)$ & $O(n_A\hat{V})$ & $O(n_A^2\hat{V}^2)$ \\
    \cline{2-5} & Ref.~\cite{Li:2024ide} & 0 & $O[\hat{V}^2\Lambda(\log_2\Lambda)^2]$ & $O[\hat{V}^2\Lambda(\log_2\Lambda)^2]$ \\
    \hline
\end{tabular}
\caption{Gate costs for implementing one Trotterized time evolution driven by the gauge field Hamiltonian, which contains the free part $\hat{H}_{\Pi}+\hat{H}_A$ and the interaction with the fermion field $\hat{H}_I$. $n_A$ scales logarithmically with the coupling $g$, the energy $\hat{E}$, the volume $\hat{V}$, and the accuracy $\epsilon$ given in Eq.~\eqref{eqn:nA_final}. The results for Ref.~\cite{Li:2024ide} are taken from Eqs.~(15) and~(16) therein by setting $d=3$ and $M=\hat{L}=\sqrt[3]{\hat{V}}$. $\Lambda$ denotes the cutoff of the photon occupation number in momentum space used in Ref.~\cite{Li:2024ide}, which scales as $\Lambda=O(g\hat{V}^3/\epsilon)+O(g^2\hat{V}^4/\epsilon)+O(\hat{V}^{7/3}/\epsilon)+O(\hat{E}\hat{V}^{4/3}/\epsilon)$ according to Eq.~(11) therein.}
\label{tab:gate_cost}
\end{table}

\section{Conclusions}
\label{sec:conclusions}
In this paper, we studied the QED Hamiltonian in the Coulomb gauge and its quantum simulation on a lattice. We first showed that in the continuum the Coulomb gauge Hamiltonian is equivalent to the temporal gauge Hamiltonian when acting on physical states. 
We then introduced a lattice discretization that maintains the decoupling of the unphysical longitudinal gauge fields and their commuting with the Hamiltonian.
The structure of the Hamiltonian guarantees that only the physical transverse gauge fields contribute to the electric and magnetic energies and thus can propagate in time evolution. Thus there is no need to impose any constraint in the simulation. We gave a
map of the gauge and fermion field basis states on the lattice onto qubits and proved that the qubit cost to represent physical states up to a given energy with a given accuracy scales polynomially with the energy, accuracy, lattice size, and Hamiltonian parameters in lattice units, see Eq.~\eqref{eqn:nA_V},
which is a new result obtained here.
Finally, we discussed a quantum algorithm to simulate the Coulomb gauge Hamiltonian with Trotterization. The gauge field basis and the conjugate variable basis at one position are swapped efficiently via the quantum Fourier transform. The fermion field operators are implemented via the Jordan-Wigner transformation. We showed that the gate costs scale polynomially with the energy, time, accuracy, lattice size, and Hamiltonian parameters in lattice units and are polynomially reduced compared with those in the previous work that uses the occupancy basis in momentum space for the gauge boson time evolution.

In future studies, one may estimate the resources needed to prepare the interacting ground state and wave packets for simulating scattering. Nevertheless, scattering is not the only interface between high energy collider physics and quantum computing. For example, one can use quantum computers to calculate parton distribution functions~\cite{Chen:2025zeh}, fragmentation functions~\cite{Grieninger:2024axp}, soft functions for jets~\cite{Bauer:2025nzf}, and energy correlators~\cite{Lee:2024jnt}.
Furthermore, one may simulate thermalization~\cite{Ebner:2024qtu} or hydrodynamization~\cite{Turro:2025sec} and extract transport coefficients~\cite{Turro:2024pxu}, which are hard problems in the field of relativistic heavy ion collisions. It is also interesting to consider the quantum simulation of non-Abelian gauge theories, e.g., the Quantum Chromodynamics in the Coulomb gauge. It is very important to understand how the Gribov copies affect real-time simulation and how to deal with them.

\acknowledgments
We would like to thank Bruno Scheihing-Hitschfeld for useful comments on the manuscript and Marc Illa Subina, Saurabh Kadam, David B. Kaplan, and Martin J. Savage for useful conversations.
This work is supported by the U.S. Department of Energy, Office of Science, Office of Nuclear Physics, InQubator for Quantum Simulation (IQuS)\footnote{\url{https://iqus.uw.edu/}} under Award Number DOE (NP) Award DE-SC0020970 via the program on Quantum Horizons\footnote{\url{https://science.osti.gov/np/Research/Quantum-Information-Science}}: QIS Research and Innovation for Nuclear Science.

\appendix
\section{Quantization Using Independent Variables}
\label{app}
Here we discuss the quantization of the continuum QED Hamiltonian in the Coulomb gauge by using independent variables. Our starting point is the QED Lagrangian density
\bea
\mathcal{L} = -\frac{1}{4}F_{\mu\nu}F^{\mu\nu} + \bar{\psi}[i\gamma^\mu (\partial_\mu - igA_\mu) - m] \psi \,,
\eea
where the field strength tensor is $F_{\mu\nu} = \partial_\mu A_\nu - \partial_\nu A_\mu$. We choose the Coulomb gauge condition $\partial_i A_i = 0$, which fixes $A_0$ to be $A_0= -\nabla^{-2} J_0$ and turns the first-class constraints into second-class 
\begin{subequations}
\bea
\partial_i A_i &=0 \,,\\
\partial_i \Pi_i &= J_0 \,,
\eea
\end{subequations}
where $\Pi_i = F_{0i}$. The conjugate variable $\Pi_{\perp i}$ introduced in Eq.~\eqref{eqn:H_coulomb} in the main text is given by $\Pi_{\perp i} = \Pi_i + \partial_iA_0$. Instead of using the Dirac quantization procedure for the constrained system, one can use independent variables and apply the standard quantization procedure with the Poisson bracket. For example, one can choose the independent variables $Q_1(\bs x) = A_1(\bs x)$ and $Q_2(\bs x) = A_2(\bs x)$ and use the constraint to write~\cite{Weinberg:1995mt}
\bea
\label{eqn:A3}
A_3(\boldsymbol x) = - \int_{-\infty}^{x_3} {\rm d}y_3 [\partial_1 Q_1(x_1,x_2,y_3) + \partial_2 Q_2(x_1,x_2,y_3) ] \,.
\eea
The canonically conjugate variables associated with $Q_1$ and $Q_2$ are
\begin{subequations}
\label{eqn:P_i}
\bea
P_1(\boldsymbol x) &= F_{01}(\boldsymbol x) + \int_{x_3}^{\infty}{\rm d}y_3 \partial_1 F_{03}(x_1,x_2,y_3) \,,\\
P_2(\boldsymbol x) &= F_{02}(\boldsymbol x) + \int_{x_3}^{\infty}{\rm d}y_3 \partial_2 F_{03}(x_1,x_2,y_3) \,.
\eea
\end{subequations}
Using integration by parts, we can show
\bea
\int {\rm d}^3x [P_1(\boldsymbol x)\dot{Q}_1(\boldsymbol x) + P_2(\boldsymbol x) \dot{Q}_2(\boldsymbol x)] = \int {\rm d}^3x [ \Pi_1(\boldsymbol x) \dot{A}_1(\boldsymbol x) + \Pi_2(\boldsymbol x) \dot{A}_2(\boldsymbol x) + \Pi_3(\boldsymbol x) \dot{A}_3(\boldsymbol x)] \,,
\eea
which just demonstrates the consistency.
The Hamiltonian density can be written as
\bea
\label{eqn:H_independent}
\mathcal{H} = P_1 \dot{Q}_1 + P_2 \dot{Q}_2 - \mathcal{L} \,,
\eea
with the commutation relations
\begin{subequations}
\bea
[Q_i(\boldsymbol x), Q_j(\boldsymbol y)] &= 0 \,,\\
[P_i(\boldsymbol x), P_j(\boldsymbol y)] &= 0 \,,\\
[Q_i(\boldsymbol x), P_j(\boldsymbol y)] &= i\delta_{ij}\delta^{(3)}(\boldsymbol x - \boldsymbol y) \,.
\eea
\end{subequations}

The last step of constructing the Hamiltonian is to express $\dot{Q}_1$ and $\dot{Q}_2$ in terms of $P_1$ and $P_2$ in Eq.~\eqref{eqn:H_independent}. This is more involved than the quantization using the constrained variables and the Dirac bracket. Using $F_{0i}=\dot{A}_i-\partial_i A_0$ in Eq.~\eqref{eqn:P_i} leads to
\begin{subequations}
\label{eqn:P_i'}
\bea
P_1(\boldsymbol x) &= \dot{A}_1(\boldsymbol x) + \int_{x_3}^{\infty}{\rm d}y_3 \partial_1 \dot{A}_3(x_1,x_2,y_3) \,,\\
P_2(\boldsymbol x) &= \dot{A}_2(\boldsymbol x) + \int_{x_3}^{\infty}{\rm d}y_3 \partial_2 \dot{A}_3(x_1,x_2,y_3) \,.
\eea
\end{subequations}
In order to invert Eq.~\eqref{eqn:P_i'}, we Fourier transform into momentum space, and regulate an integral as
\bea
\int_{x_3}^{\infty}{\rm d}y_3 \int_{-\infty}^{y_3} {\rm d}z_3 e^{ip_3z_3} \to \lim_{\epsilon\to0}\int_{x_3}^{\infty}{\rm d}y_3 \int_{-\infty}^{y_3} {\rm d}z_3 \, e^{ip_3z_3-\epsilon |z_3|} = \frac{e^{ip_3x_3}}{p_3^2} \,,
\eea
which is valid as long as $p_3\neq 0$. Then we obtain
\bea
\begin{bmatrix}
1+\frac{p_1^2}{p_3^2} & \frac{p_1p_2}{p_3^2} \\
\frac{p_1p_2}{p_3^2} & 1+\frac{p_2^2}{p_3^2}
\end{bmatrix}
\begin{bmatrix}
\dot{Q}_1(\bs p) \\
\dot{Q}_2(\bs p)
\end{bmatrix} = 
\begin{bmatrix}
P_1(\bs p)\\
P_2(\bs p)
\end{bmatrix} \,,
\eea
which can be solved as long as $\bs p\neq0$.

Finally, we explicitly write $\mathcal{H}$ as
\bea
\mathcal{H} = P_1\dot{Q}_1 + P_2\dot{Q}_2 - \frac{1}{2}(\dot{Q}_1^2 + \dot{Q}_2^2 + \dot{A}_3^2) - J^1Q_1-J^2Q_2-J^3A_3 - \frac{1}{2}J^0A_0 - \bar{\psi}(i\gamma^i\partial_i - m)\psi \,,
\eea
where $A_3$ is a function of $Q_1$ and $Q_2$ as in Eq.~\eqref{eqn:A3} and dotted variables are written in terms of $P_1$ and $P_2$ as
\begin{subequations}
\bea
\dot{Q}_1(\boldsymbol x) &= \frac{\partial_2^2+\partial_3^2}{\nabla^2} P_1(\boldsymbol x) - \frac{\partial_1\partial_2}{\nabla^2} P_2(\boldsymbol x) \,, \\
\dot{Q}_2(\boldsymbol x) &= -\frac{\partial_1\partial_2}{\nabla^2} P_1(\boldsymbol x) + \frac{\partial_1^2+\partial_3^2}{\nabla^2} P_2(\boldsymbol x) \,, \\
\dot{A}_3(\boldsymbol x) &= - \int_{-\infty}^{x_3} {\rm d}y_3 [\partial_1 \dot{Q}_1(x_1,x_2,y_3) + \partial_2 \dot{Q}_2(x_1,x_2,y_3) ] \,.
\eea
\end{subequations}
In the Hamiltonian $H=\int {\rm d}^3x \mathcal{H}(\bs x)$, the $\dot{Q}_i^2$ terms involve triple volume integrals and the $\dot{A}_3^2$ term involves triple volume integrals plus two line integrals, which are numerically more expensive to implement in lattice calculations, compared with the $\Pi_{\perp i}^2$ term in the Coulomb gauge Hamiltonian in terms of the constrained variables, which only contains double volume integrals, see Eq.~\eqref{eqn:H_E_coulomb}.

\bibliography{main}
\end{document}